\documentclass[manuscript]{aastex}

\slugcomment{Accepted: Astrophysical Journal Suppl., Feb. 2009}

\shorttitle{Ring Galaxy Atlas}
\shortauthors{Madore, Nelson \& Petrillo}

\begin{document}


\title{Atlas and Catalog of Collisional Ring Galaxies}


\author{\bf Barry F. Madore}
\affil{Observatories of the  Carnegie Institution of Washington \\ 813 Santa 
Barbara St., Pasadena, CA ~~91101}
\author{\bf Erica Nelson}
\affil{Observatories of the  Carnegie Institution of Washington \\ 813 Santa 
Barbara St., Pasadena, CA ~~91101\\
and
\\Pomona College\\
Claremont, CA ~91711\\
and
\\Department of Astronomy\\
Yale University\\
J.W. Gibbs Laboratory \\
260 Whiteney Ave.\\
New Haven, CT ~06511}
\author{\bf  Kristen Petrillo}
\affil{Observatories of the  Carnegie Institution of Washington \\ 813 Santa 
Barbara St., Pasadena, CA ~~91101\\
and
\\Pomona College\\
Claremont, CA ~91711}

\email{barry@ociw.edu, erica.nelson@pomona.edu, kristen.petrillo@pomona.edu}



\begin{abstract}
We present a catalog and imaging atlas of classical (collisional) {\tt
RING} galaxies distilled from the Arp-Madore {\it Atlas of Southern
Peculiar Galaxies and Associations} and supplemented with other known
{\tt RING} galaxies from the published literature. The catalog lists
the original host object, compiles available redshifts and presents
newly determined positions for the central (target) galaxy and its
nearest companion(s). 127 collisional {\tt RING} systems are
illustrated and their components identified.  All of the {\tt RINGS}
have plausible colliders identified; many are radial-velocity
confirmed companions.  Finally, we make note of the existence of a
rare sub-class of {\tt RING} galaxies exemplified by AM 2136-492,
double/concentric {\tt RING} galaxies.  These objects are predicted by
numerical simulations, but they appear to be quite rare and/or
short-lived in nature.

\end{abstract}

\vfill\eject
\section{INTRODUCTION}
Arp \& Madore (1987, hereafter AM87) undertook a fairly comprehensive
survey of the southern sky in search of peculiar and interacting
galaxies visible on the IIIa-J photographic survey plates as part of
the UK Schmidt sky survey. They inspected some 100,000 galaxies and
culled out about 7,000 objects that were noteworthy enough to be
cataloged, and in a number of cases, individually illustrated.  Based
on this statistically robust sample AM87 went beyond simple cataloging
and added a broad list of descriptors covering about two dozen
properties. One of these descriptor/classifications was Category 6:
Ring Galaxy.

As already noted by AM87 ``Ring Galaxy'', in the isolated context of
this particular catalog, was a purely morphologically defined
descriptor; as a consequence this descriptor encompasses several
physically distinct types of galaxies. First, there are relatively
normal galaxies with ``outer rings''.  In the RC3 system of de
Vaulcouleurs et al. these are typed as {\bf RS} galaxies, and called
``outer ringed'' galaxies in the compilation by Buta (1995).  Outer
rings are not uncommon and are generally well understood (see Buta for
an overview). Second, there are ``polar ring galaxies''; systems
thought to be advanced mergers with the disrupted satellite now
orbiting the parent galaxy over the pole and forming the ring-like
structure. These are exceedingly rare objects (see Whitmore et
al. 1990 for a recent compilation and references therein.) Finally,
there are the collisional systems (called ``{\tt RING}'' in RC3); and
they are the focus of this paper.

In an earlier paper (which was based on an incomplete, early-release
version of the Arp-Madore Catalogue) Few, Madore \& Arp (1986)
(hereafter FMA) pointed out the two dominant types of rings in the
catalog and separated them into two distinct classes: {\bf O-Types}
for the outer (``resonant'') rings, and {\bf P-Types} for the
classical (``collisional'') rings. The main conclusion of that paper
was that the {\bf P-type} ({\tt RING}) galaxies were consistent with
the head-on collision scenario, being championed in that same time
period by Theys \& Spiegel (1976) and independently by Lynds \& Toomre
(1976). The full success of that theory has broader implications which
are explored in Madore et al. (2009).

\section{The Catalog}

All three of the current authors independently inspected and
reclassified each of the 500+ Category~6 (Ring Galaxies) in AM87. Only
{\bf P-type} galaxies (on the FMA classification system) were retained for
this study. That is, any object with a crisp, relatively
high-surface-brightness ring, with the nucleus often asymmetrically
placed (or in some cases absent), was given the P classification and
retained. Objects with more diffuse or feathery outer rings (of
generally somewhat lower surface brightness), with fairly large
nuclear regions or bulges symmetrically placed with respect to the
ring, were deemed to be the {\bf O Types}. These objects were returned to
the general pool. Out of 552 Class~6 (Ring) objects in AM87 we culled
out 104 {\bf P-Type} {\tt RING} Galaxies; and for only a handful of objects
was there any disagreement among the authors as to their
classification. Erring on the conservative/inclusive side, all of the
contentious objects were retained in this study.

Table 1 lists all 104 pure {\bf P-Type} ({\tt RING}) galaxies found in AM87.
We note here that pure {\tt RING} galaxies constitute less than 2\% of
the AM87 peculiar galaxy population, and as such they are found in the
general population of galaxies only 0.01\% of the time (as normalized
by the entire population of $\sim$100,000 galaxies inspected by AM87 in the
course of constructing the catalog as a whole.) {\tt RING} galaxies
are a rare type of collision with short-lived evidence.  It is of
course interesting to contemplate what {\tt RING} galaxies look like
after the collisionally-induced burst of star formation fades.  We
leave that as an exercise for the future.

Rather than simply illustrating a few of the most spectacular examples
of {\tt RING} galaxies, as was done in AM87, we have chosen to
re-catalog and present the entire sample. However, before doing that
we decided to additionally survey the literature for previously
cataloged {\tt RING} galaxies and add them to the {\it Atlas} for
completeness sake. The AM87 sample is given in Table 1, while the {\tt
RING} galaxies from the literature are collected in Table 2. Clearly
the northern and equatorial sky is incompletely sampled. We estimate
that about 200 more {\tt RING} galaxies await discovery once those
regions are inspected to the same apparent-size and surface-brightness
limit as AM87.

Table 1 and 2 share the same basic formating. Column 1 contains the
name of the host {\tt RING} galaxy complex; the second column contains
the name or names of the ring nucleus followed by as many companions
as are suspected of being colliders. The name of the ring nucleus is
composed of the name of the host object followed by RN (e.g.,
AM~0147-350:RN); by way of contrast, companion/colliders are
identified by the host object name followed by C1, C2, etc. (e.g.,
AM~0147-350:C1, AM~0147-350:C2, AM~0147-350:C3). Columns 4 and 5
respectively contain RA(2000) and Dec(2000) positions, each measured
to a precision of 5~arcsec. In Column 5 the first entry (D =
diameter), corresponds to the diameter of the ring, measured in
arcsec, while the following entries in that same column but now for
the companions are marked by S, and give the separation of the
companion from the nucleus of the ring (again, measured in
arcsec). Finally, Column 6 gives the heliocentric velocity for the
object/component when available from the literature as captured by NED
in its December 2008 release.

\section{The Ring Galaxy Sample}

\subsection{Some General Remarks and  Demographics}

All of the objects in this new listing consist of at least two objects
in the immediate field of view that can reasonably be identified with
the collider and the (collidee) ring. We mark those objects in each of
the illustrations and for uniformity sake we have measured the
positions of all objects given in Tables 1 and 2 using the display
capabilities of ALADIN interacting with the NED image archive. The
presence of nearby companions was by no means a foregone conclusion
given that the {\tt RINGS} were chosen purely on the basis of their
morphology, not on the basis of their nearby or distant
environment. We consider it highly likely that these optical
companions are the colliders, plausibly responsible for the
``gravitational splash'' and the ensuing ring-like, star-formation
response predicted/post-dicted by the simulations. Indeed, of the 64
companions with published redshifts (found from a search of the
January 2008 release of the NED database) 60 have redshifts that place
them at the same distance as the ring itself. In all four of the other
instances where one of the apparent companions proved to be a
background object there was always an additional candidate collider in
the same field of view. Although common radial velocities are not
conclusive proof that any given object is the collider it does clearly
demonstrate that the vast majority of those objects are not chance
line-of-sight background (or foreground) objects. 111 apparent
companions still need redshifts; 43 of the rings themselves are
without published velocities. We are working to ameliorate that
situation.

While all of our {\tt RING} galaxies have one or more apparent
companions, as a class {\tt RING} galaxies tend not to be found in
high-density (cluster) environments. Indeed, only 7 of the 132 {\tt RINGS}
cataloged here have more than three companions in their immediate
vicinity. They are AM~0401-641, AM~0417-391, AM~0455-465, AM~1003-215,
AM~1251-283, AM~2100-725 and AM~2200-715.  83 of the {\tt RINGS} have only
one nearby optical companion, 29 have two companions, and only 13 have
three. It seems reasonable to suppose that in dense systems the more
frequent, but off-center encounters with other group members will
disrupt and/or destroy any ring that might form from a direct hit more
quickly than in the quiescent field. Perhaps the increased velocity
dispersion of all the members associated with the cluster environment
also plays a role.  And it should not be forgotten that very rich
clusters are {\it currently} depleted of disk galaxies in any case:
with no disks to be hit, no rings can be formed. However, at high
redshift the situation is expected to be quite different.  A search
for {\tt RINGS} in various ``deep fields'' (UDF, GOODS and GEMS) has
been made (Elmegreen \& Elmegreen 2006). According to those authors
and judging from the published images of their pure ``rings'' (their
Figure 1) these are almost exclusively {\bf O-Type} (outer-ring)
galaxies with well-centered bulges and no obvious companions.  Some
collisional objects may be found in their sample of fifteen ``partial
rings''; however, the area shown around each partial ring is too small
to judge whether these objects have interacting companions or not.

Figure 1 shows the wide range of absolute (K-band) magnitudes that the
{\tt RING} galaxies in this catalog have. Spanning more than a factor
of 100 in luminosity galaxies with ring-like morphology can be as
faint as M$_K$ = -19~mag and as bright as M$_K$ = -25~mag; typical
{\tt RING} galaxies have M$_K$ = -23~mag which is somewhat fainter
than the knee in the general-field luminosity function. The colliders
are typically half a magnitude fainter than the central (target)
galaxy (Figure 2), but their dispersion is large ($\pm$0.75~mag).  The
current diameters of the rings cataloged here span a wide range being
broadly peaked at 30-40~kpc, but extending up to values as large as
70~kpc (Figure 3). None of these plots argue for separate populations
of {\tt RING} galaxies based solely on these physical properies.

\subsection{Empty Rings}

Some {\tt RING} galaxies (e.g., AM 0058-220, AM 1953-260, Arp~147 \&
VII~Zw~466) do not have any obvious component that can unambiguously be
identified as a nucleus. The morphology of such objects has been aptly
likened to ``smoke rings.'' The apparent lack of a nucleus may be
because the target galaxy simply never had a particularly noteworthy
nucleus to begin with (i.e., even before the collision) or it may be
that the timing of the interaction, combined with our particular
viewing angle places the nucleus in the ring (see NGC~4774, Arp~146)
and/or even confused with the intruder.  In other instances the
nucleus may have been disrupted by the collision.  Detailed modeling
would illuminate these possibilities.

\subsection{Double Rings}

We draw attention to a sub-class of objects in this compilation of
{\tt RING} galaxies: double rings.  AM~2136-492 is noteworthy. The
Cartwheel (AM~0035-335) is another example (with its celebrated
compact nuclear ring); while  AM~0339-625, AM~1323-222 and AM~1354-250
may be other less obvious examples of this class. Figure 6 shows a
deeply stretched image of  AM~0339-625 illustrating its fragmented,
outer ring-like feature.  Although multiple rings structures seem to
be a common feature in the numerical simulations (e.g. Antunes \&
Wallin 2007) they appears not to be not so common in nature.

\section{Implications}
A simple calculation concerning the apparent frequency of {\tt RING}
galaxies leads to a rather surprising/interesting result. The relative
frequency of the collisonal ring phenomenon is at least 1 in 1000,
given that AM87 inspected approximately 100,000 galaxies and out of
those cataloged only 100 rings. (And this is a lower limit given that
edge-on rings are probably under-represented in the catalog.) If the
ring structure can unambiguously be detected as such for one dynamical
timescale ($10^8$ yrs, say) and if the collision rate is constant over
cosmic time, then about one in every ten galaxies in this same volume
is expected to have been perturbed in this same way over a Hubble
time.  Given the restricted parameter space available for a head-on
collision to occur this strongly suggests that, if the orbits of these
types of companions are isotropic (and only a small fraction of them
produce rings) then the majority of galaxies in the local volume have
presumably been hit (with a different impact parameter) by a companion
comparable in mass to that of a typical ring-galaxy collider.

In Figure 4 we show the histogram of relative velocities from those
{\tt RING} galaxies with redshift measurements for both the ring
itself and at least one of its putative colliders. The median velocity
difference for this sample is rather small ($\sim$100~km/s). Excluding
obvious background objects with $\Delta$V $>$ 1,000~km/s, the
largest two remaining velocity differences are 293 and 419~km/s. We
conclude that the majority of these systems are bound and that if
simulations are a guide then there will be a rapid merging of these
pairs in the course of the next passage.

\section{Discussion \& Conclusions}

Computer simulations of galaxy interactions first showed that {\tt
RING} galaxies could convinciungly be created by the head-on collision
of a satellite galaxy with a primary disk. Observations of
prototypical {\tt RING} galaxies, such as the Cartwheel (Fosbury \&
Hawarden 1977) and the Lindsay-Shapley ring, each confirmed the
``prediction'' in as much as colliders were found in evidence, and
within the range of distances, velocities and position angles expected
for head-on collisional events.  The present study follows the early
work of Few, Madore \& Arp (1986) and extends this confirmation of
theory from a handful of galaxies to a relative complete sample of
over 100 suspected and previously cataloged {\tt RING} galaxies.

However, there is another prediction that is fundamentally at odds
with what is observed for this sample of {\tt RING} galaxies. That
prediction comes from $\Lambda$-CDM cosmology in which many more
satellite galaxies are expected to be orbiting central galaxies than
are in fact observed. Making these satellites non-luminous (be it by
supernova events, by interactions, by initial conditions, or by fiat)
does not prevent them from interacting gravitationally, and as such
they too should be responsible for creating rings through collisions
in direct proportion to their numbers with respect to their luminous
colliding counterparts.  The astonishing fact is that {\it there is
not a single compelling example of a {\tt RING} galaxy without a
plausible optical collider.} There is no isolated Cartwheel in the
sky; there are no smoke rings without the smoking gun.  And there
should be many. We state this simple fact and leave a more detailed
discussion of this point and its implications for $\Lambda$-CDM
cosmology to the companion paper by Madore et al. (2009).

\centerline{\it Acknowledgements} We wish to acknowledge and thank the
creators of SAOImage DS9 and Aladin both of which were used
extensively in the preparation of this {\it Atlas}.  And, of course,
this research has made use of the NASA/IPAC Extragalactic Database
(NED) which is operated by the Jet Propulsion Laboratory, California
Institute of Technology, under contract with the National Aeronautics
and Space Administration. Finally, we thank Ian Steer for bringing the
ring galaxy IC~1908 to our attention.

\noindent
\centerline{\bf References \rm}
\vskip 0.1cm
\vskip 0.1cm

\par\noindent 
Antunes, A. \& Wallin, J. 2007, ApJ, 670, 261

\par\noindent 
Arp, H.C. 1966, ApJS, 14, 1

\par\noindent 
Arp, H.C. \& Madore, B.F. 1987 {\it Catalogue of
Southern Peculiar Galaxies and Associations}, Cambridge University
Press, Cambridge.

\par\noindent 
Buta, R. 1995, ApJS, 96, 39

\par\noindent 
Buta, R. \& Combes, F. 1996, Fund. Cosmic Physics, 17, 9

\par\noindent 
Elmegreen, D.M. \& Elmegreen, B.G. 2006, ApJ, 651, 676

\par\noindent 
Few, J.M.A., Arp, H.C. \& Madore, B.F.  1982, MNRAS, 199, 633

\par\noindent 
Few, J.M.A., Madore, B.F. \& Arp, H.C., 1986, MNRAS, 222, 673

\par\noindent 
Fosbury, R.A.E. \& Hawarden, T.G. 1977, MNRAS, 178, 473

\par\noindent 
Lynds, C.R. \& Toomre, A. 1976, ApJ, 209, 382

\par\noindent 
Madore, B.F., Nelson, E. \&  Petrillo, K. 2009, ApJ, (submitted)

\par\noindent 
Marston, A.P. \& Appleton, P.N. 1995, AJ, 109, 1002

\par\noindent 
Whitmore, B.C., Lucas, R., McElroy, D.B., Steinman-Cameron, T., 
Sackett, P.D. \& Oiling, R.P. 1990, AJ, 100, 1489

\par\noindent 
Theys, J.C. \& Spiegel, E.A. 1976, ApJ, 208, 650

\vfill\eject
\vskip 0.75cm


\clearpage
\begin{figure}
\begin{center}
\includegraphics[angle=-90, scale=0.70]{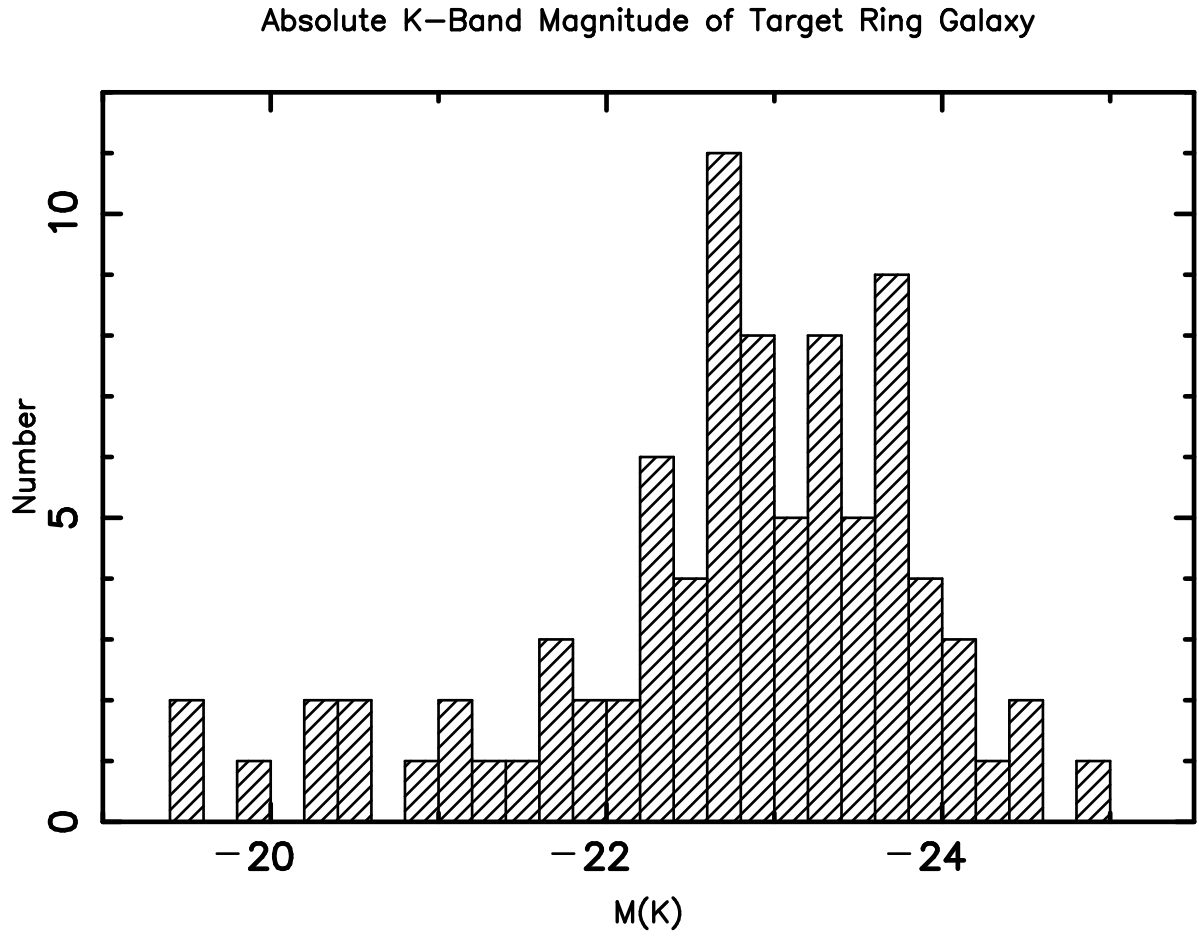}
\caption{Histogram of the absolute K-band magnitudes of the central ({\tt RING}) galaxy.}
\end{center}
\end{figure}

\clearpage
\begin{figure}
\begin{center}
\includegraphics[angle=-90, scale=0.70]{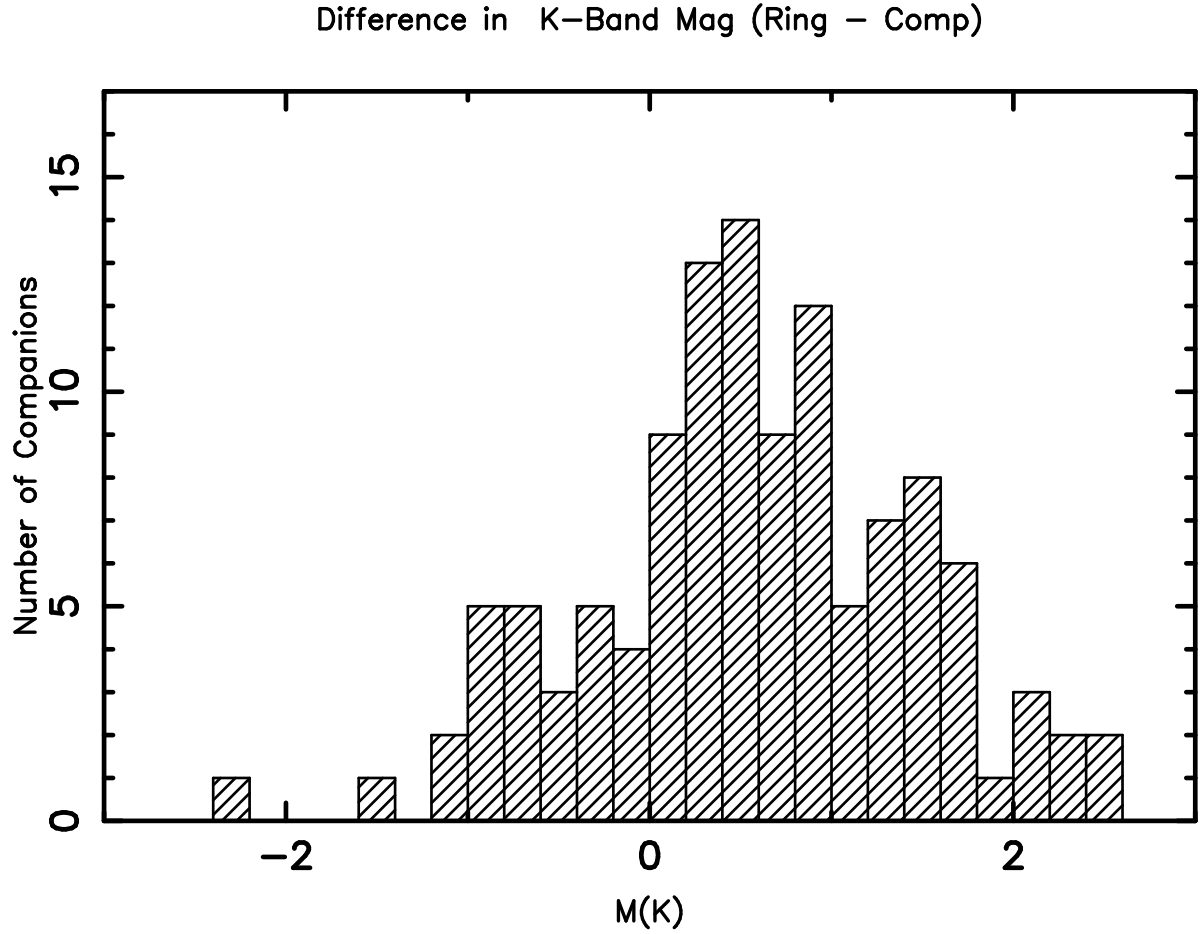}
\caption{Histogram of the K-band magnitude differences between the central {\tt RING} and its collider.}
\end{center}
\end{figure}

\clearpage
\begin{figure}
\begin{center}
\includegraphics[angle=-90, scale=0.70]{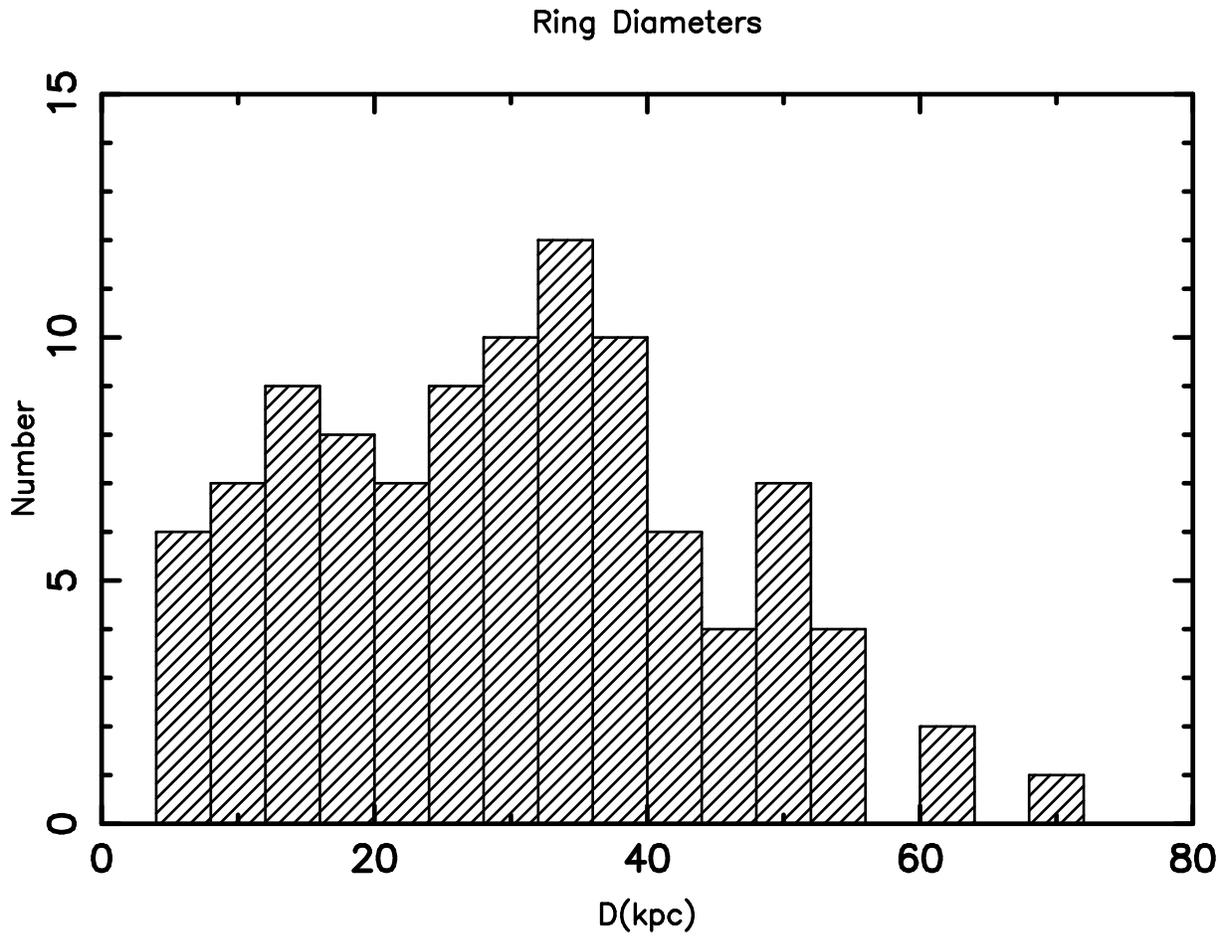}
\caption{Histogram of the metric major-axis diameters of the rings in units of kpc.}
\end{center}
\end{figure}

\clearpage
\begin{figure}
\begin{center}
\includegraphics[angle=-90, scale=0.70]{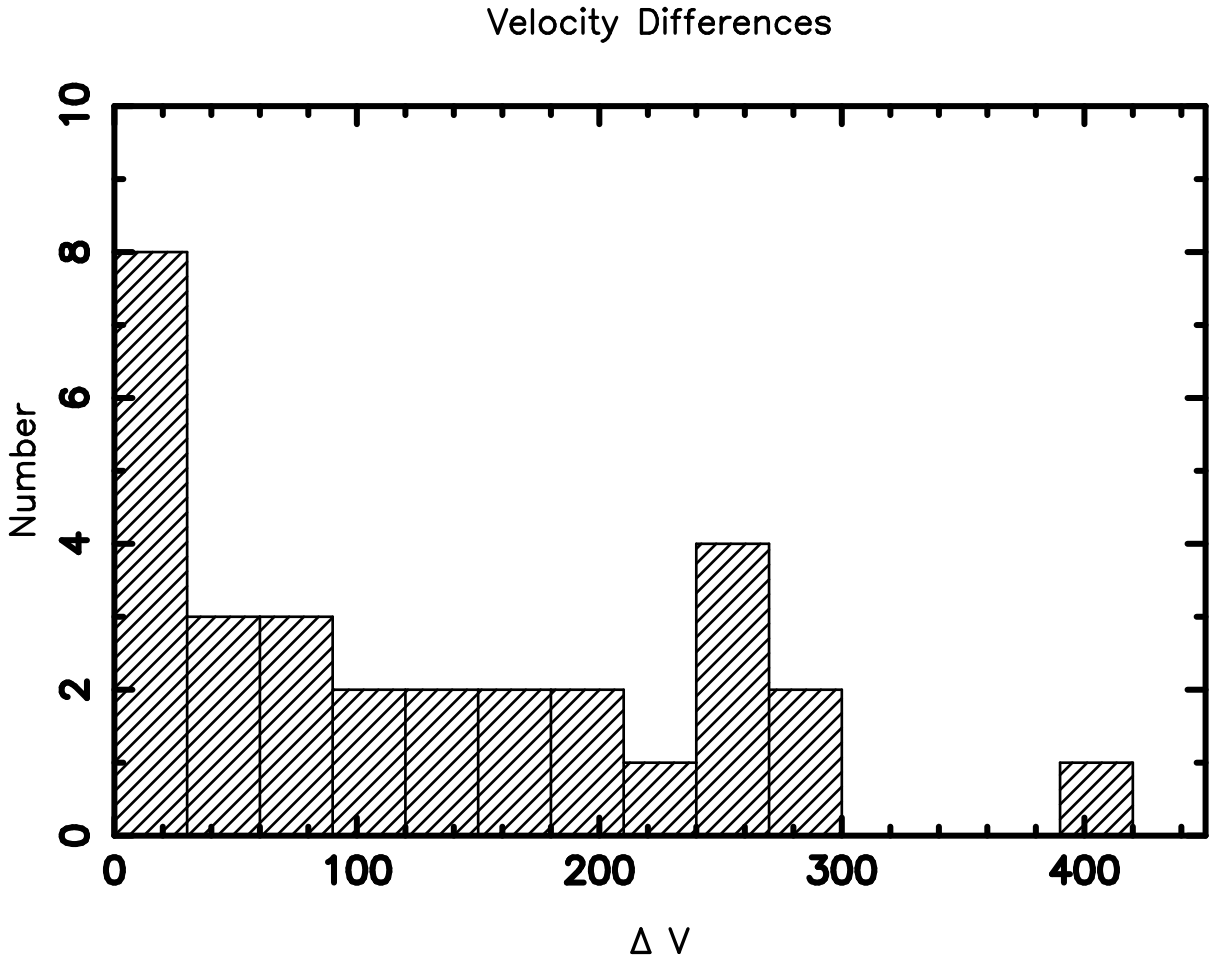}
\caption{Histogram of the velocity differences between the ring and its collider.}
\end{center}
\end{figure}


\begin{figure}
\resizebox{0.51\hsize}{!}{\includegraphics[width=0.4\textwidth]{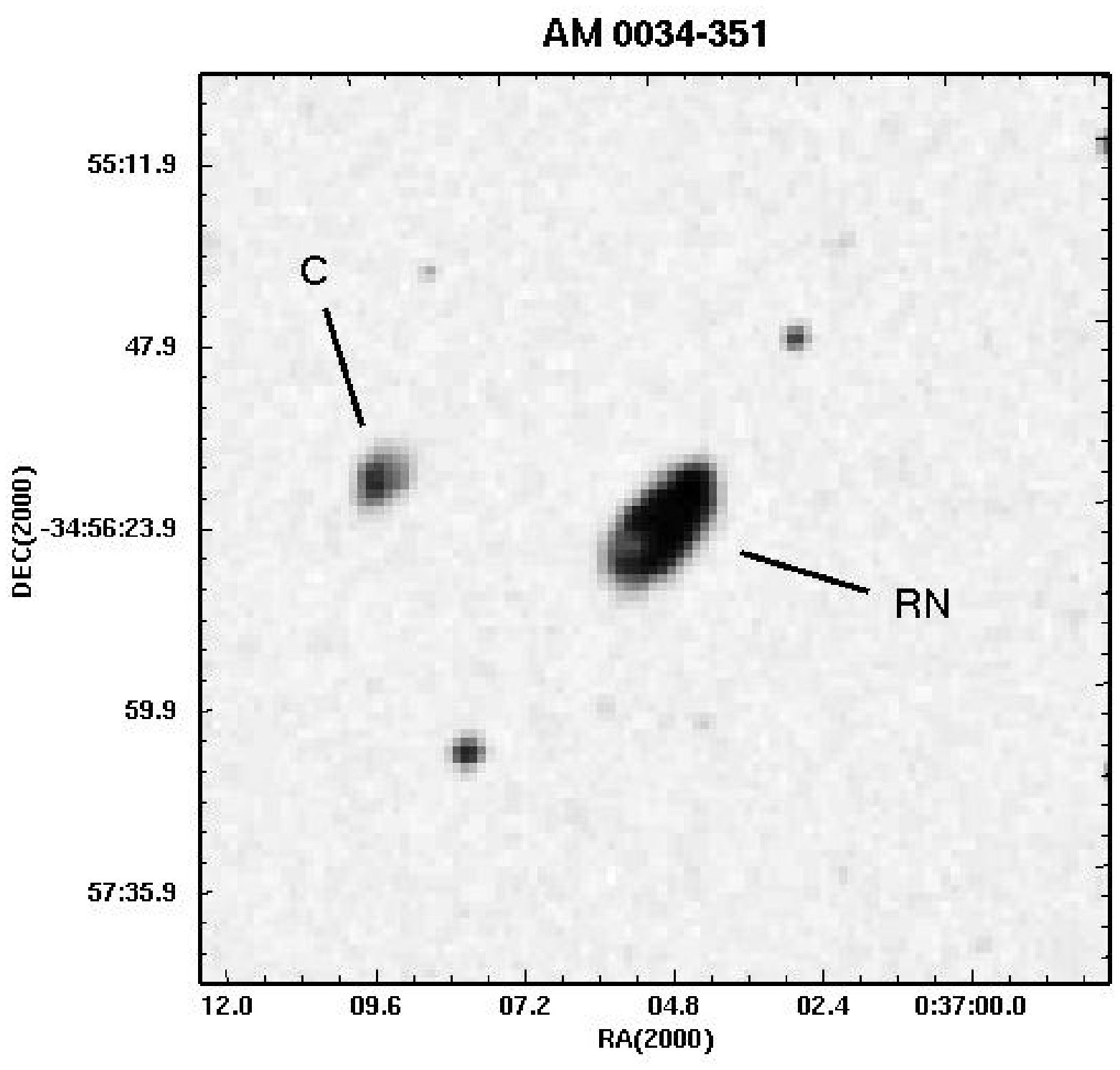}}\hspace{0.02cm} 
\resizebox{0.55\hsize}{!}{\includegraphics{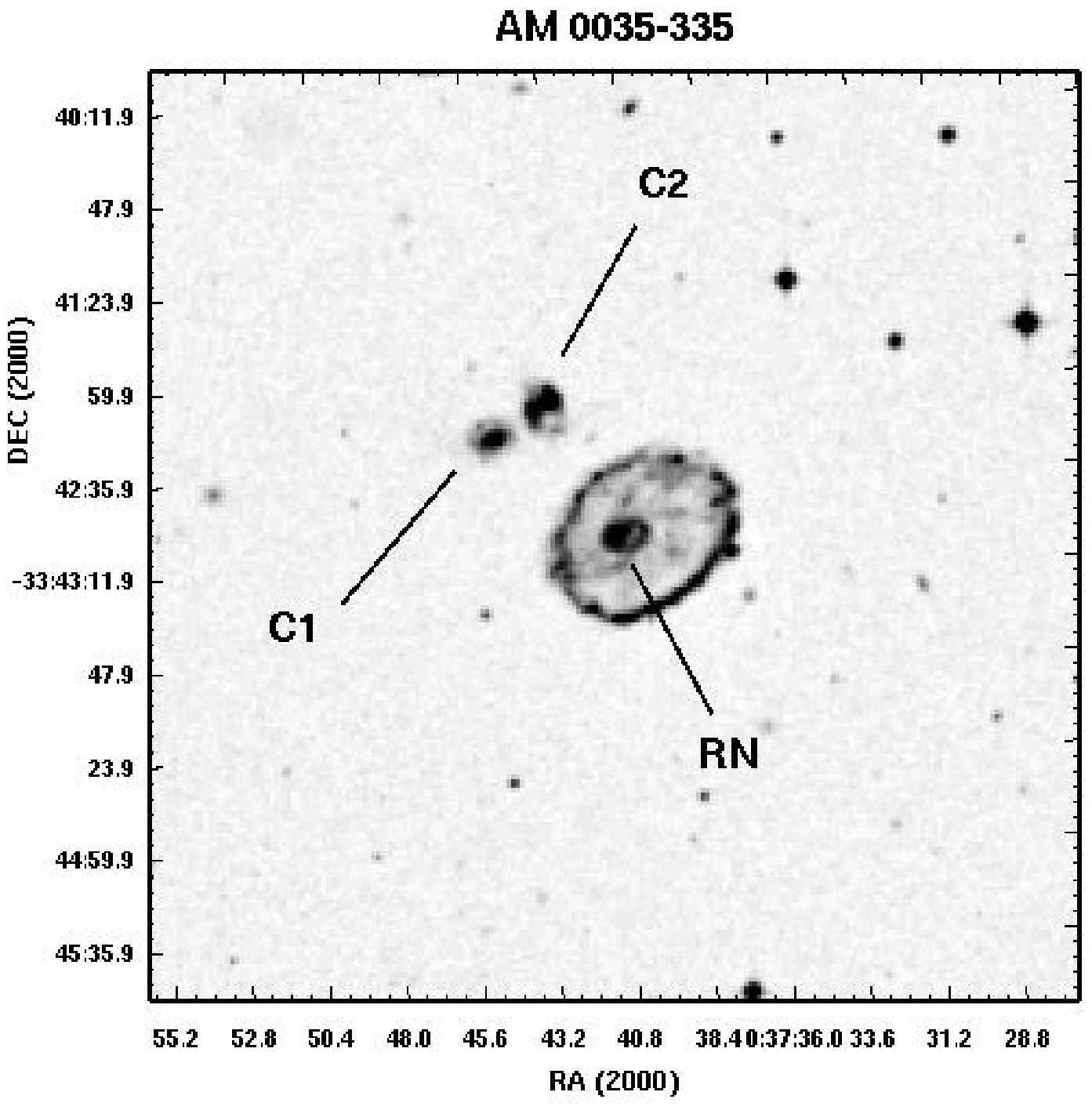}}\\

\resizebox{0.55\hsize}{!}{\includegraphics{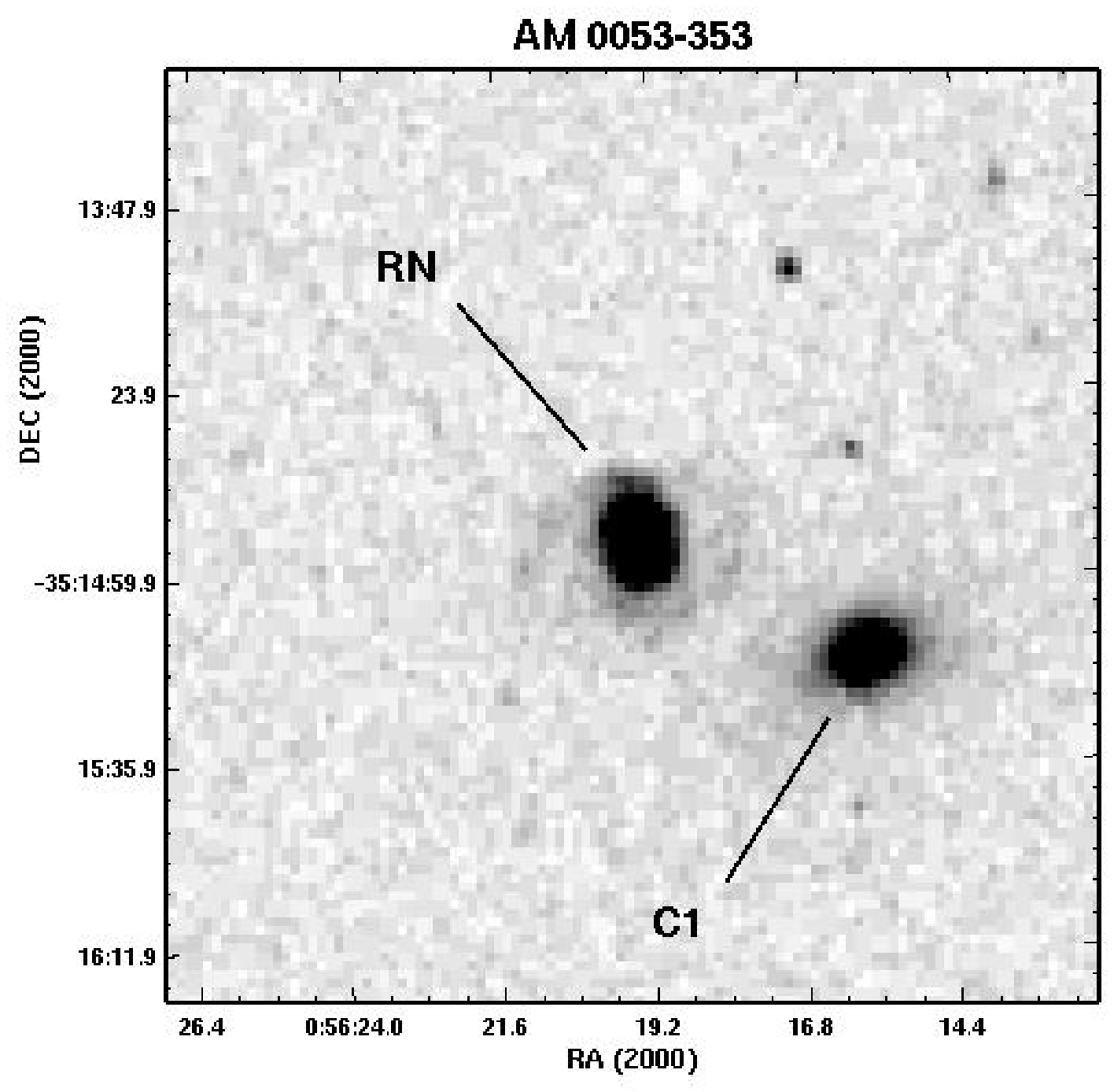}}\hspace{0.1cm}
\resizebox{0.51\hsize}{!}{\includegraphics{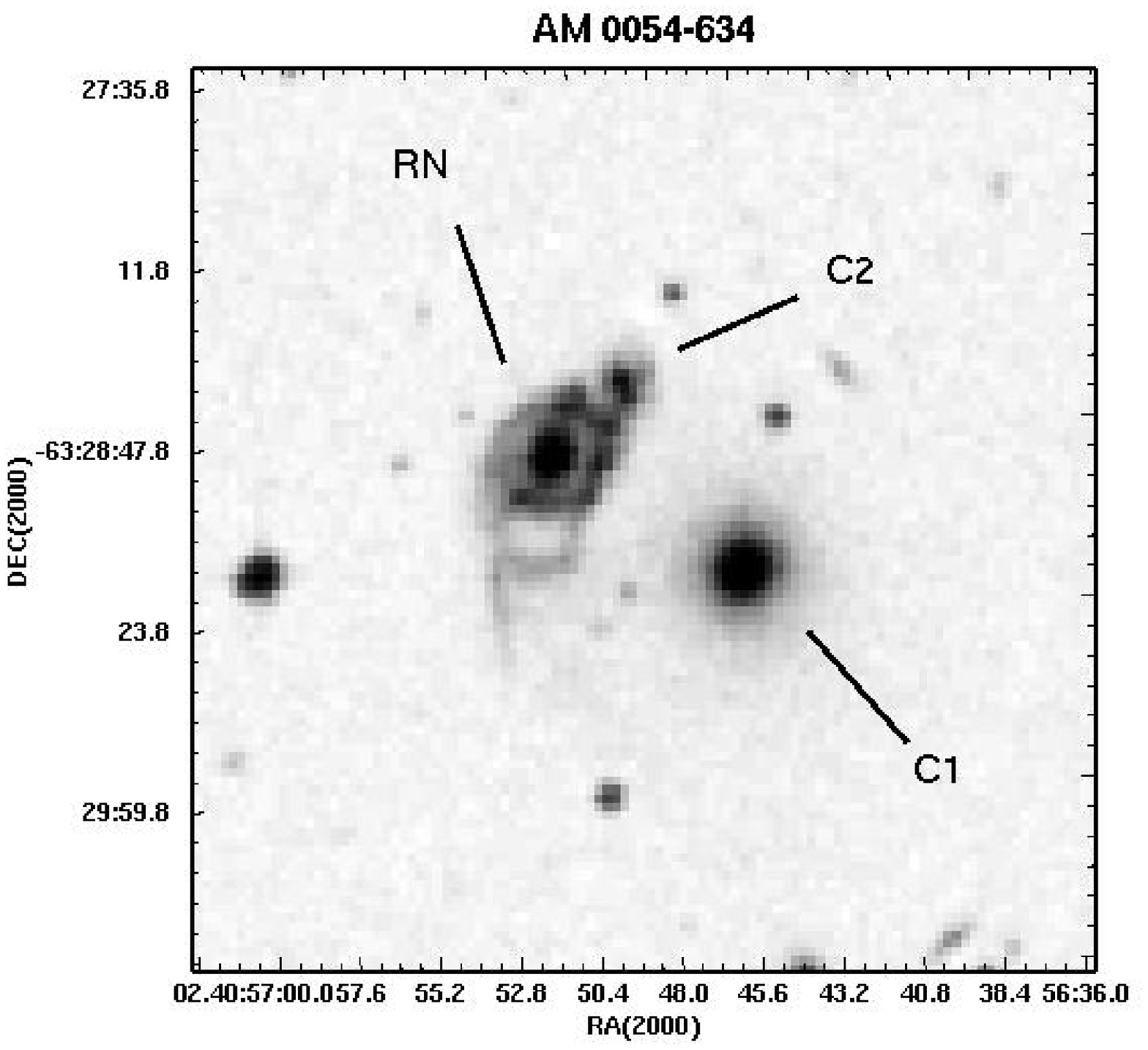}}\\

\resizebox{0.51\hsize}{!}{\includegraphics{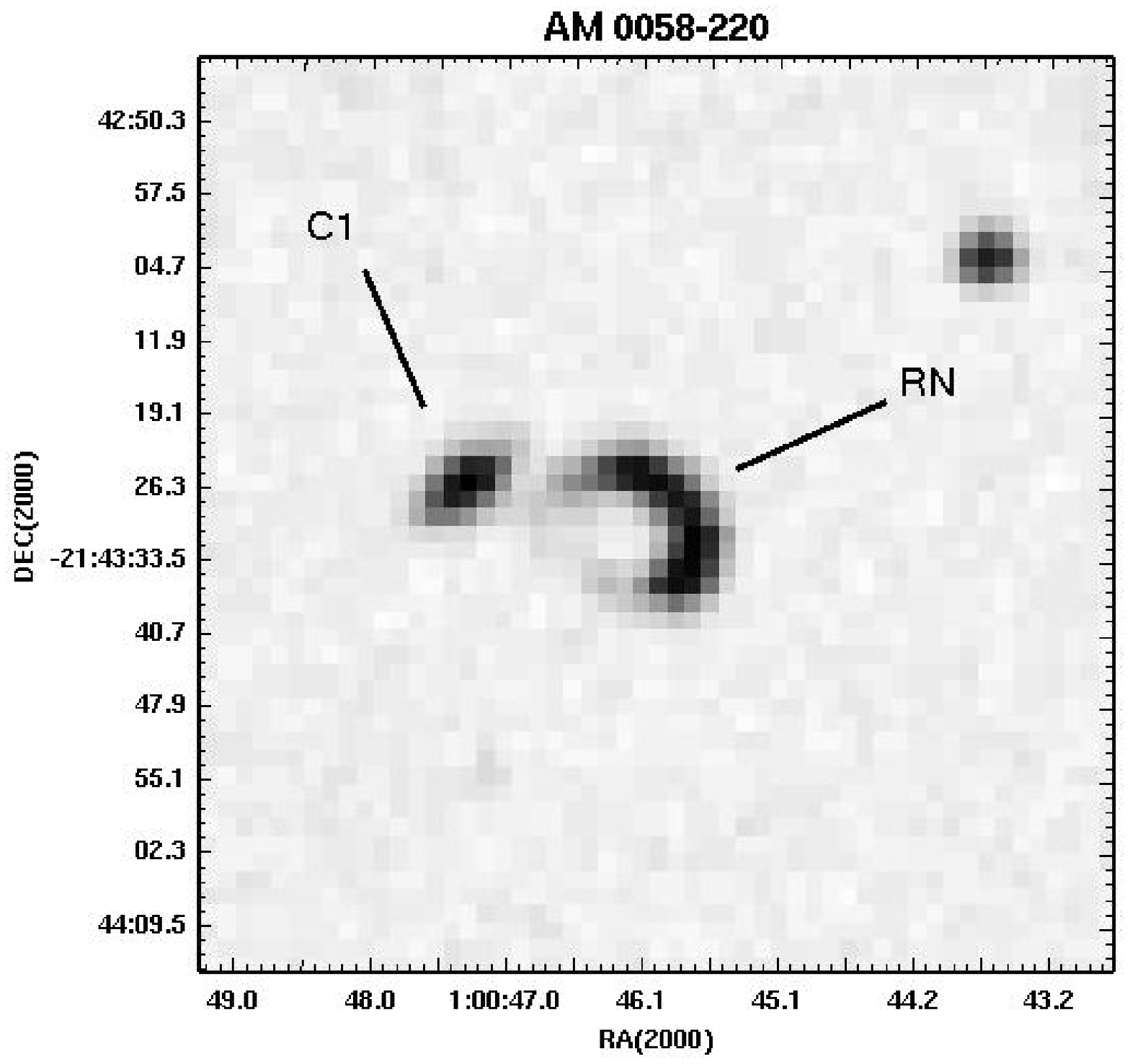}}\hspace{0.1cm}
\resizebox{0.51\hsize}{!}{\includegraphics{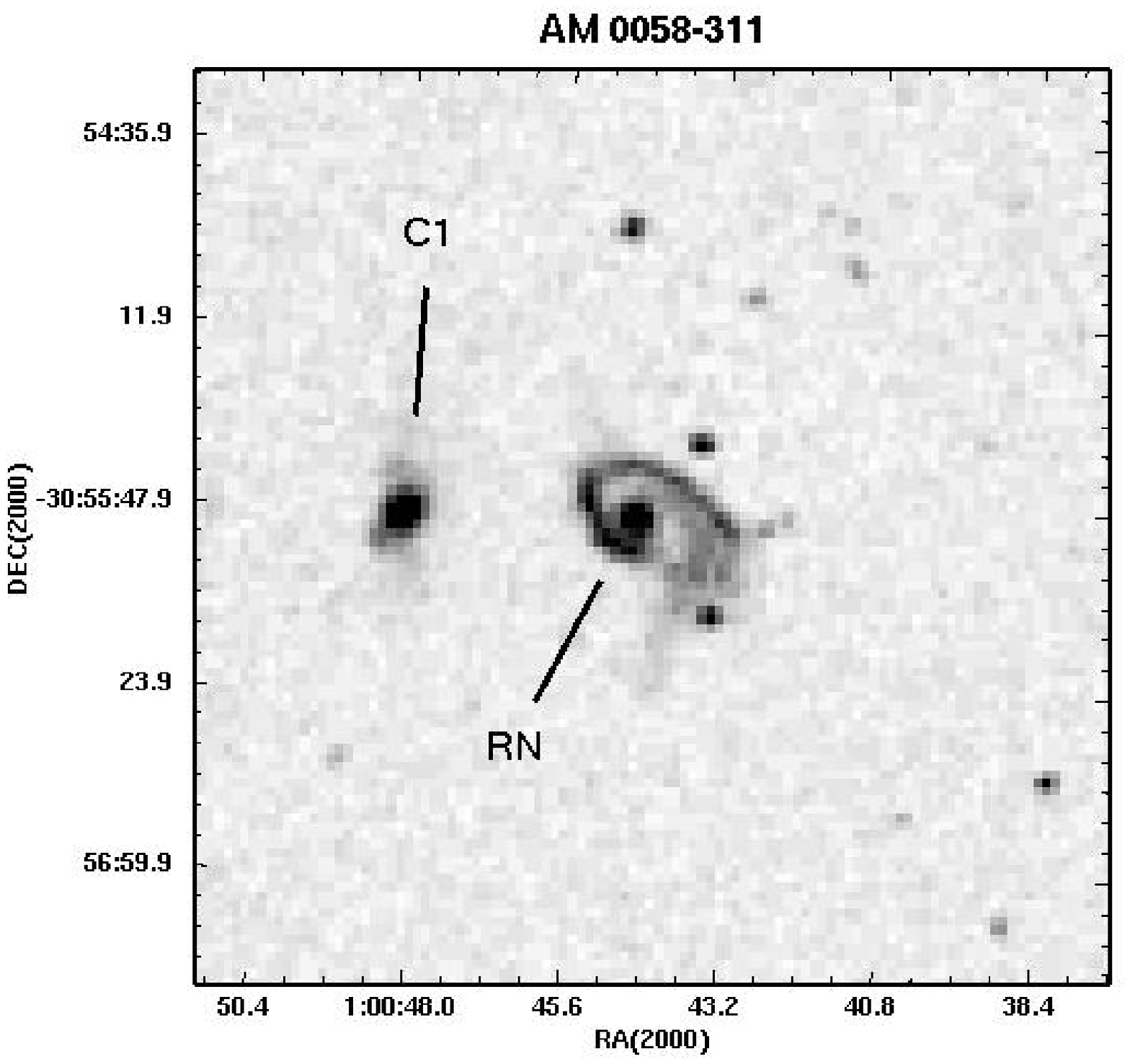}}\\
\caption{Ring Galaxy Atlas}
\end{figure}

\begin{figure}
\resizebox{0.51\hsize}{!}{\includegraphics{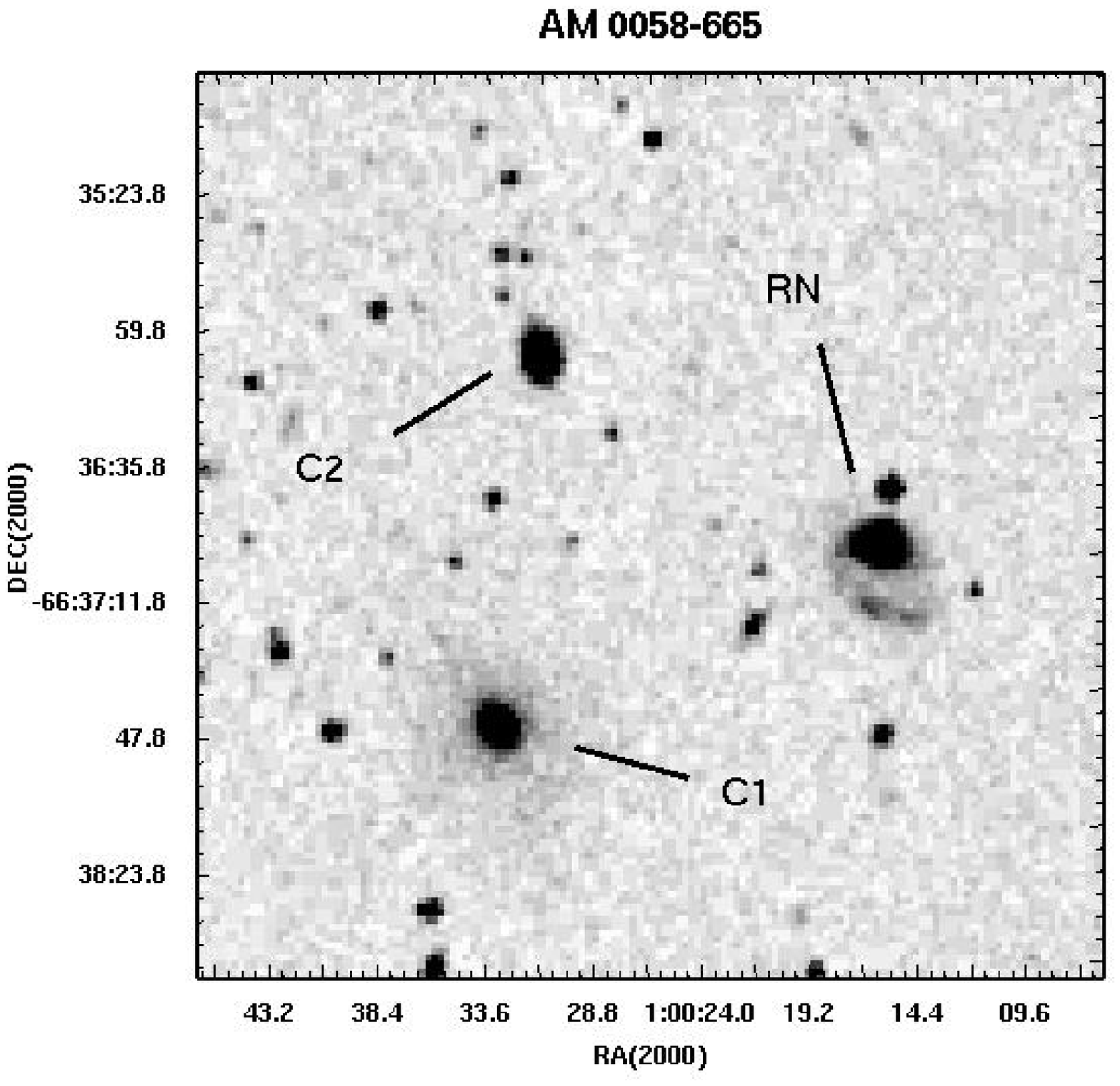}}\hspace{0.02cm} 
\resizebox{0.51\hsize}{!}{\includegraphics{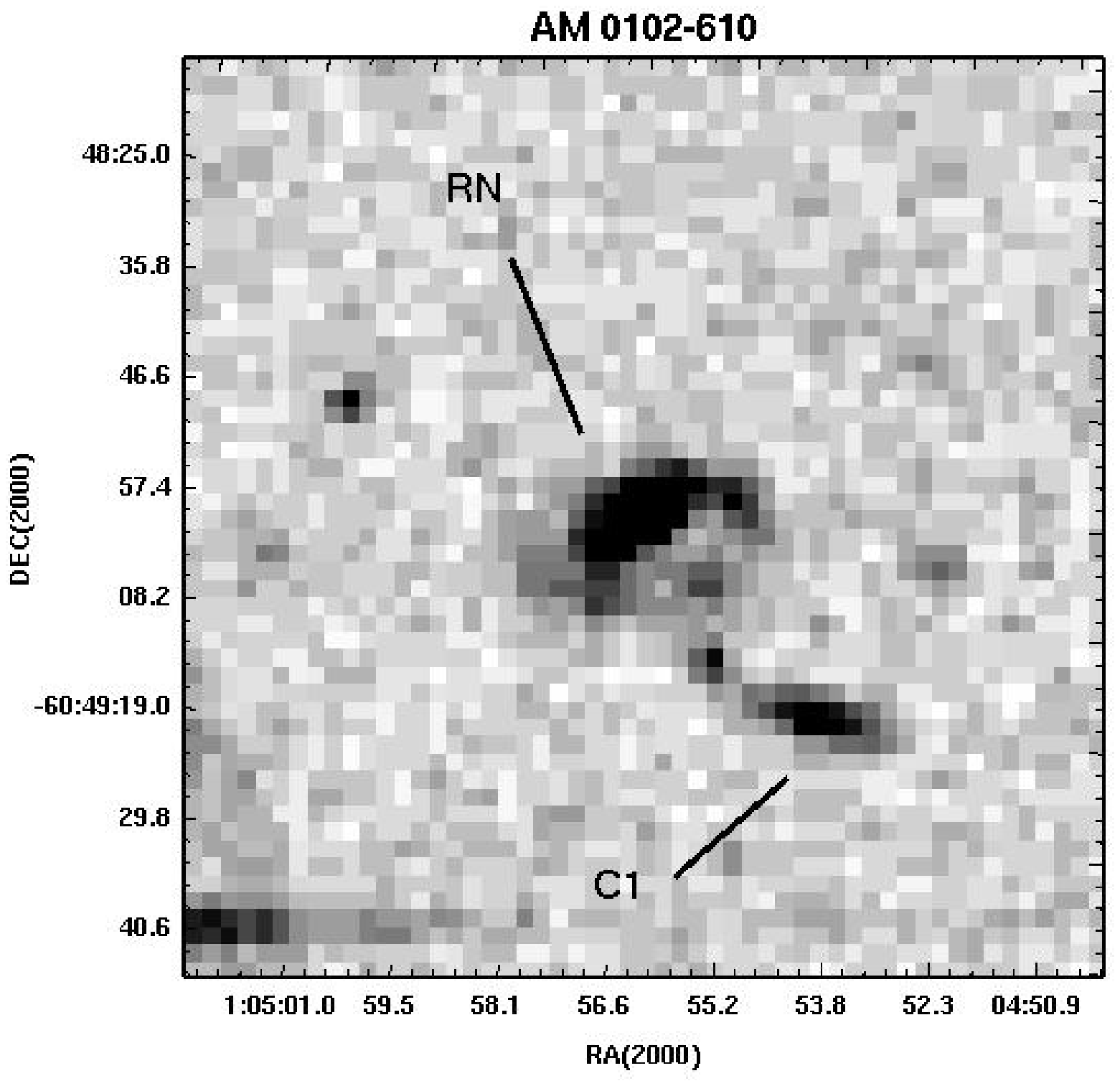}}\\

\resizebox{0.51\hsize}{!}{\includegraphics{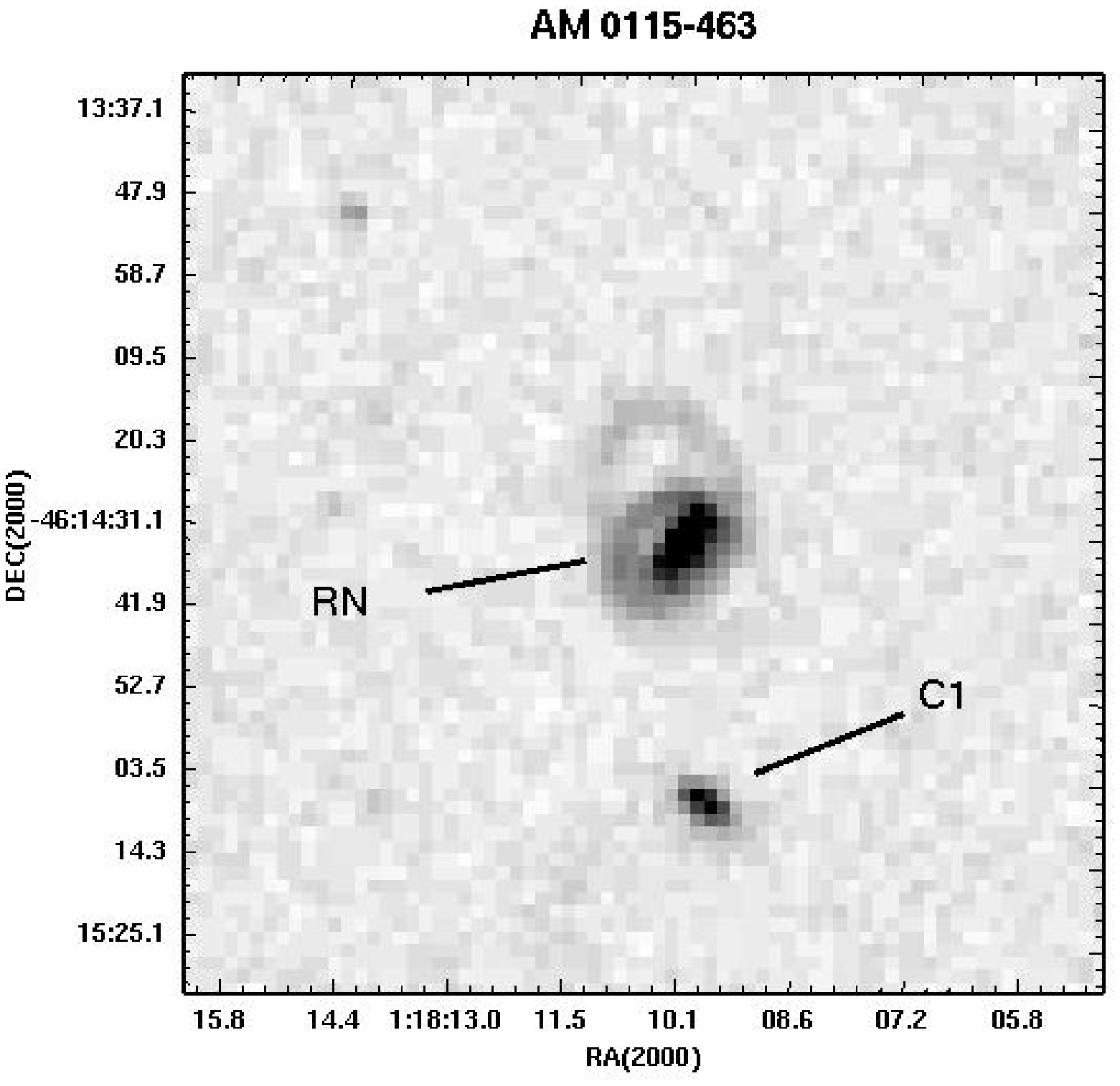}}\hspace{0.1cm}
\resizebox{0.51\hsize}{!}{\includegraphics{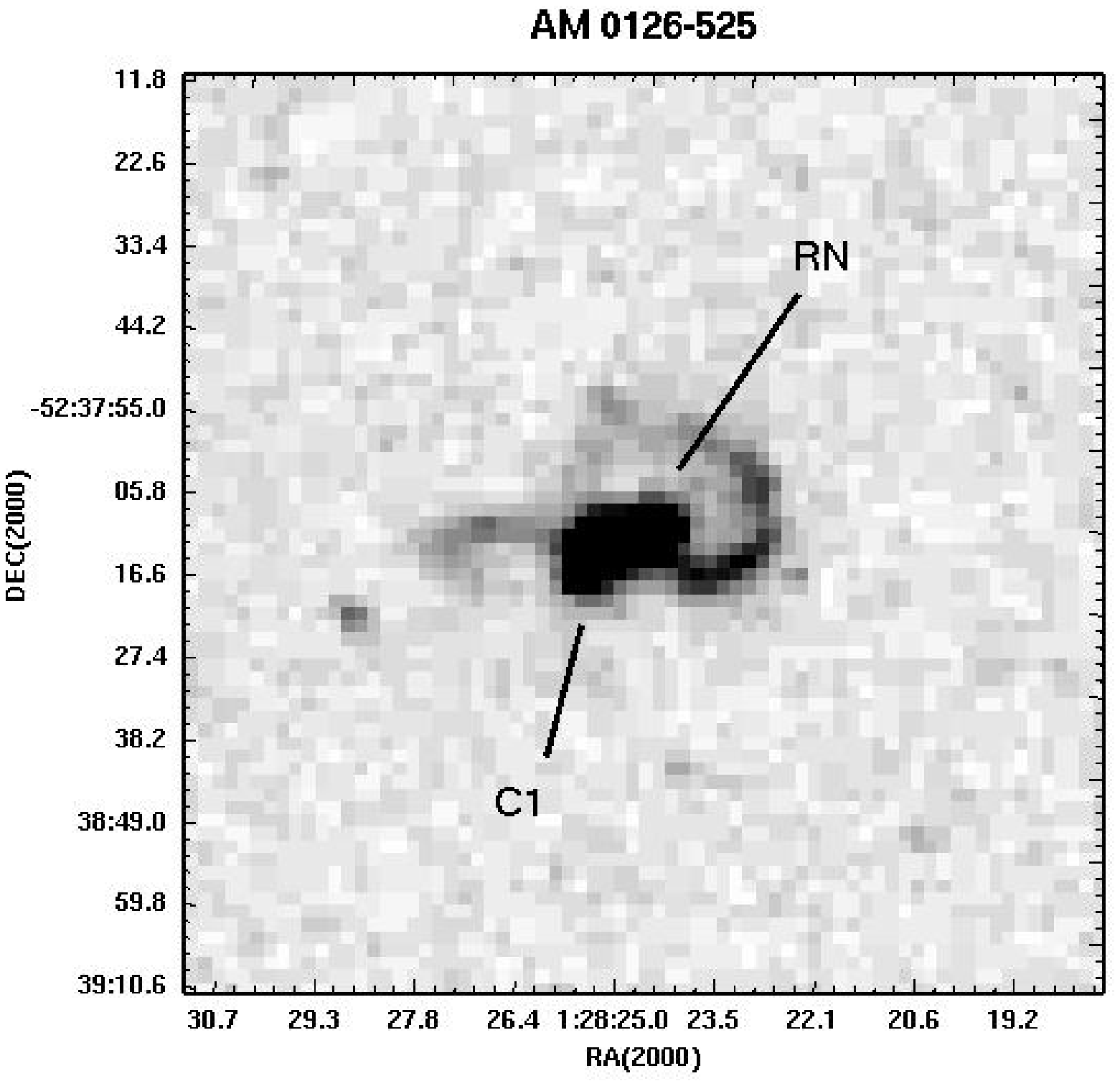}}\\

\resizebox{0.51\hsize}{!}{\includegraphics{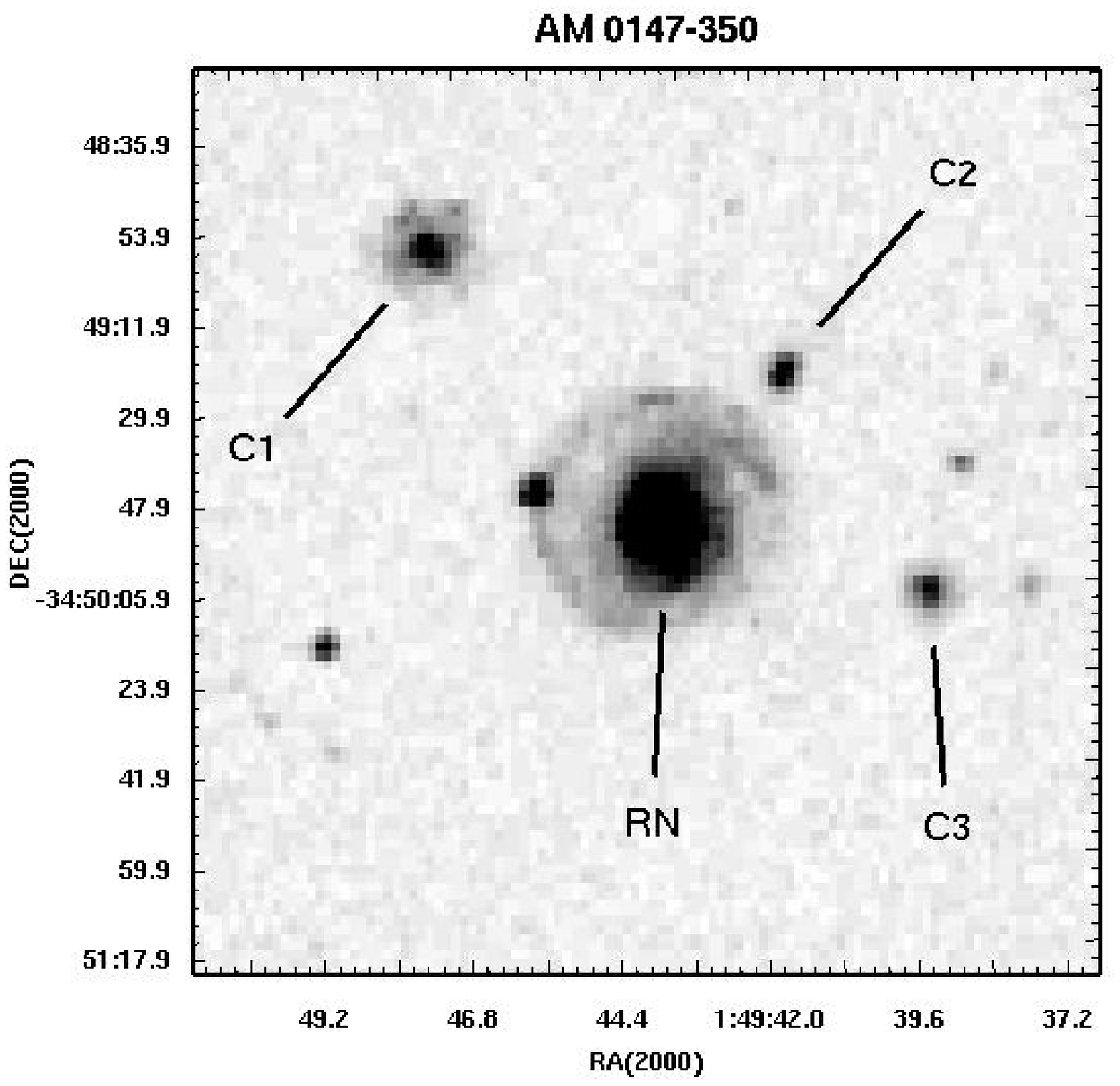}}\hspace{0.1cm}
\resizebox{0.51\hsize}{!}{\includegraphics{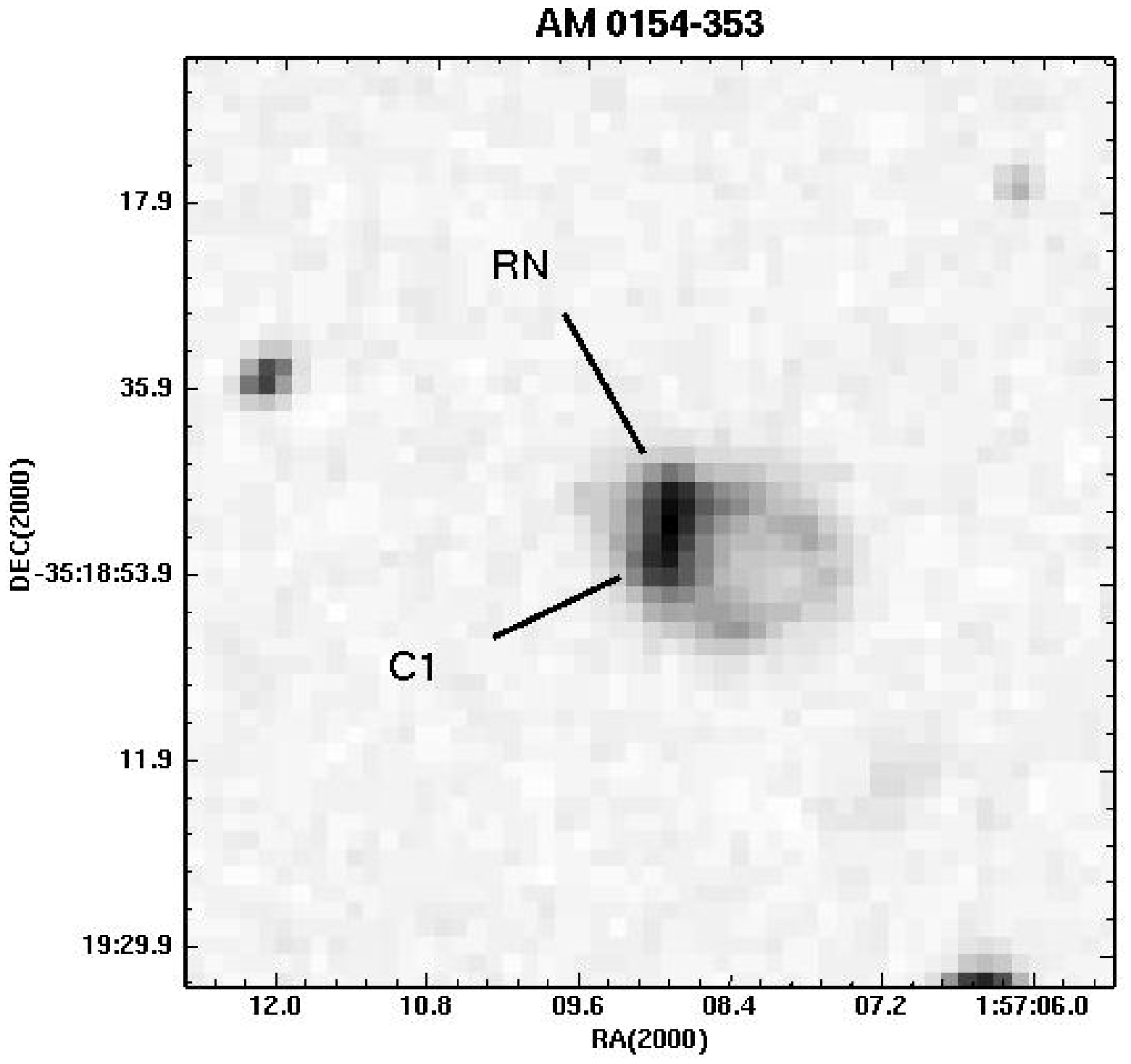}}\\
\end{figure}

\vfill\eject
\noindent

\begin{deluxetable}{llcclr}
\tablecolumns{5}
\tablewidth{6.6truein}
\tablehead{
\colhead{Host Ring System}  & \colhead{Component}  & \colhead{RA (2000)}    & \colhead{Dec (2000)}  & \colhead{~~D/S} &\colhead{V$_{\odot}$}\\
\colhead{}    & \colhead{Name}    & \colhead{(h:m:s)} & \colhead{(d:m:s)}   & \colhead{~~(arcsec)} &\colhead{(km/s)} 
}

\startdata
AM 0034-351&AM 0034-351:RN & 00:37:09.3& -34:56:15 & D = ~31 & 8,891\\
           &AM 0034-351:C1 & 00:37:04.7& -34:56:24 & S = ~58 & . . .~~  \\
& & & & & \\
AM 0035-335&AM 0035-335:RN & 00:37:41.1& -33:42:59 & D = ~58 & 9,050\\
           &AM 0035-335:C1 & 00:37:44.7& -33:42:20 & S = ~61 & 9,104\\
           &AM 0035-335:C2 & 00:37:43.1& -33:42:08 & S = ~60 & 8,639\\
& & & & & \\
AM 0053-353&AM 0053-353:RN & 00:56:19.3& -35:14:53 & D = ~47 & 14,494\\
           &AM 0053-353:C1 & 00:56:15.7& -35:15:15 & S = ~43 & 14,420\\
& & & & & \\
AM 0054-634&AM 0054-634:RN & 00:56:51.1& -63:28:51 & D = ~27 & 11,557\\
           &AM 0054-634:C1 & 00:56:45.5& -63:29:16 & S = ~20 & 11,526\\
           &AM 0054-634:C2 & 00:56:48.8& -63:28:38 & S = ~45 & . . .~~  \\
& & & & & \\
AM 0058-220&AM 0058-220:RN & 01:00:46.0& -21:43:30 & D = ~17 & . . .~~  \\
           &AM 0058-220:C1 & 01:00:47.2& -21:43:26 & S = ~17 & 12,524\\
& & & & \\
AM 0058-311&AM 0058-311:RN & 01:00:44.5& -33:55:49 & D = ~34 & 23,650\\
           &AM 0058-311:C1 & 01:00:48.1& -33:55:49 & S = ~46 & 23,684\\
& & & & & \\
AM 0058-665&AM 0058-665:RN & 01:00:15.1& -66:37:07 & D = ~36 & 21,055\\
           &AM 0058-665:C1 & 01:00:32.3& -66:37:49 & S = 112 & . . .~~  \\
           &AM 0058-665:C2 & 01:00:29.4& -66:36:11 & S = 103 & . . .~~  \\
& & & & & \\
AM 0102-610&AM 0102-610:RN & 01:04:56.3& -60:48:58 & D = ~22 & . . .~~  \\
           &AM 0102-610:C1 & 01:04:53.8& -60:49:17 & S = ~26 & . . .~~  \\
& & & & & \\
AM 0115-463&AM 0115-463:RN & 01:18:09.9& -46:14:32 & D = ~33 & 17,688\\
           &AM 0115-463:C1 & 01:18:09.7& -46:15:07 & S = ~34 & . . .~~  \\
& & & & & \\
AM 0126-525&AM 0126-525:RN & 01:28:24.6& -52:38:08 & D = ~29 & 3,700\\
           &AM 0126-525:C1 & 01:28:25.2& -52:38:11 & S = ~06 & 3,450\\
& & & & & \\
AM 0147-350&AM 0147-350:RN & 01:49:43.5& -34:49:53 & D = ~52 & 8,244\\
           &AM 0147-350:C1 & 01:49:47.2& -34:48:57 & S = ~73 & . . .~~\\
           &AM 0147-350:C2 & 01:49:41.4& -34:49:25 & S = ~39 & . . .~~\\
           &AM 0147-350:C3 & 01:49:39.2& -34:50:08 & S = ~55 & bckgrnd\\
& & & & & \\
AM 0154-353&AM 0154-353:RN & 01:57:08.8& -35:18:47 & D = ~23 & . . .~~\\
           &AM 0154-353:C1 & 01:57:08.9& -35:18:51 & S = ~06 & . . .~~\\

AM 0155-815&AM 0155-815:RN & 01:54:41.4& -81:37:53 & D = ~38 & . . .~~\\
           &AM 0155-815:C1 & 01:53:00.8& -81:37:57 & S = ~43 & . . .~~\\
           &AM 0155-815:C2 & 01:52:39.7& -81:37:50 & S = ~05 & . . .~~\\
& & & & & \\
AM 0157-311&AM 0157-311:RN & 01:59:39.5& -31:01:11 & D = ~29 & 17,742\\
           &AM 0157-311:C1 & 01:59:37.8& -31:01:22 & S = ~24 & . . .~~\\
& & & & & \\
AM 0200-581&AM 0200-581:RN & 02:01:50.8& -57:56:52 & D = ~27 & . . .~~\\
           &AM 0200-581:C1 & 02:01:50.7& -57:57:01 & S = ~10 & . . .~~\\
           &AM 0200-581:C2 & 02:01:51.4& -57:56:56 & S = ~06 & . . .~~\\
& & & & & \\
AM 0240-242&AM 0240-242:RN & 02:42:32.0& -24:13:52 & D = ~26 & 25,310\\
           &AM 0240-242:C1 & 02:42:32.0& -24:13:57 & S = ~05 & . . .~~\\
           &AM 0240-242:C2 & 02:42:33.4& -24:13:05 & S = ~46 & . . .~~\\
& & & & & \\
AM 0305-824&AM 0305-824:RN & 03:01:44.3& -82:35:30 & D = ~34 & . . .~~\\
           &AM 0305-824:C1 & 03:02:01.8& -82:35:26 & S = ~34 & . . .~~\\
& & & & & \\
AM 0318-610&AM 0318-610:RN & 03:19:37.5& -60:57:01 & D = ~17 & . . .~~\\
           &AM 0318-610:C1 & 03:19:37.9& -60:57:10 & S = ~09 & . . .~~\\
AM 0322-374&AM 0322-374:RN & 03:23:54.6& -37:30:36 & D = ~43 & 1,405\\
           &AM 0322-374:C1 & 03:23:53.7& -37:30:47 & S = ~15 & 1,397\\
& & & & & \\
AM 0339-625&AM 0339-625:RN & 03:40:37.8& -62:46:49 & D = 30/62 & 14,860\\
           &AM 0339-625:C1 & 03:40:39.6& -62:46:36 & S = ~18 & . . .~~\\
& & & & & \\
AM 0401-641&AM 0401-641:RN & 04:01:36.9& -64:04:03 & D = ~40 & . . .~~\\
           &AM 0401-641:C1 & 04:01:43.4& -64:05:29 & S = ~16 & . . .~~\\
           &AM 0401-641:C2 & 04:01:54.9& -64:06:10 & S = ~85 & 12,457 \\
           &AM 0401-641:C3 & 04:01:37.2& -64:04:38 & S = ~78 & . . .~~\\
           &AM 0401-641:C4 & 04:01:46.7& -64:05:54 & S = ~28 & . . .~~\\
& & & & & \\
AM 0403-555&AM 0403-555:RN & 04:04:16.4& -55:46:23 & D = ~40 & . . .~~\\
           &AM 0403-555:C1 & 04:04:17.0& -55:46:06 & S = ~18 & 17,825\\
           &AM 0403-555:C2 & 04:04:12.1& -55:45:20 & S = ~80 & . . .~~\\
& & & & & \\
AM 0404-271&AM 0404-271:RN & 04:06:28.8& -27:09:21 & D = ~25 & 26,443\\
           &AM 0404-271:C1 & 04:06:29.4& -27:09:24 & S = ~08 & . . .~~\\
& & & & & \\
& & & & & \\
AM 0413-590&AM 0413-590:RN & 04:14:25.9& -58:57:45 & D = ~43 & 17,143\\
           &AM 0413-590:C1 & 04:14:27.0& -58:57:40 & S = ~09 & . . .~~\\
& & & & & \\
AM 0415-475&AM 0415-475:RN & 04:16:29.9& -47:50:54 & D gt  84 & 10,079\\
           &AM 0415-475:C1 & 04:16:18.6& -47:49:12 & S = 152 & . . .~~\\
           &AM 0415-475:C2 & 04:16:31.6& -47:50:43 & S = ~23 & . . .~~\\
& & & & & \\
AM 0417-391&AM 0417-391:RN & 04:19:39.6& -39:10:31 & D = ~31 & 15,255\\
           &AM 0417-391:C1 & 04:19:38.9& -39:10:26 & S = ~09 & . . .~~\\
           &AM 0417-391:C2 & 04:19:39.9& -39:10:23 & S = ~10 & 15,255\\
           &AM 0417-391:C3 & 04:19:40.5& -39:08:40 & S = 114 & 14,820 \\
           &AM 0417-391:C4 & 04:19:40.4& -39:11:39 & S = ~70 & . . .~~\\
           &AM 0417-391:C5 & 04:19:51.3& -39:10:05 & S = 140 & 15,185\\
& & & & & \\
AM 0425-421&AM 0425-421:RN & 04:26:36.6& -42:05:38 & D = ~55 & 4,491\\
           &AM 0425-421:C1 & 04:26:44.6& -42:05:42 & S = ~91 & . . .~~\\
& & & & & \\
AM 0437-314&AM 0437-314:RN & 04:39:40.0& -31:42:39 & D = ~28 & . . .~~\\
           &AM 0437-314:C1 & 04:39:40.3& -31:42:38 & S = ~05 & . . .~~\\
& & & & & \\
AM 0437-644&AM 0437-644:RN & 04:37:54.9& -64:38:59 & D = ~14 & . . .~~\\
           &AM 0437-644:C1 & 04:37:56.9& -64:39:06 & S = ~33 & . . .~~\\
& & & & & \\
AM 0438-503&AM 0438-503:RN & 04:39:20.5& -50:31:50 & D = ~38 & . . .~~\\
           &AM 0438-503:C1 & 04:39:22.1& -50:31:45 & S = ~15 & . . .~~\\
& & & & & \\
AM 0438-661&AM 0438-661:RN & 04:38:37.1& -66:13:03 & D = ~34 & 14,650\\
           &AM 0438-661:C1 & 04:38:37.9& -66:13:49 & D = ~12 & . . .~~\\
& & & & & \\
AM 0455-465&AM 0455-465:RN & 04:57:01.7& -46:45:00 & D = ~15 & . . .~~\\
           &AM 0455-465:C1 & 04:57:04.1& -46:44:54 & S = ~26 & . . .~~\\
           &AM 0455-465:C2 & 04:57:06.5& -46:45:03 & S = ~49 & . . .~~\\
           &AM 0455-465:C3 & 04:57:00.1& -46:46:00 & S = ~60 & . . .~~\\
           &AM 0455-465:C4 & 04:56:55.8& -46:44:10 & S = ~83 & . . .~~\\
           &AM 0455-465:C5 & 04:57:06.5& -46:46:08 & S = ~80 & . . .~~\\
& & & & & \\
AM 0507-512&AM 0507-512:RN & 05:08:28.0& -51:21:29 & D = ~34 & . . .~~\\
           &AM 0507-512:C1 & 05:08:27.6& -51:20:38 & S = ~43 & . . .~~\\
           &AM 0507-512:C2 & 05:08:27.3& -51:21:22 & S = ~11 & . . .~~\\
           &AM 0507-512:C3 & 05:08:29.5& -51:21:42 & S = ~20 & . . .~~\\

AM 0520-390&AM 0520-390:RN & 05:22:42.5& -39:03:48 & D = ~43 & 15,114\\
           &AM 0520-390:C1 & 05:22:43.4& -39:03:25 & S = ~26 & . . .~~\\
           &AM 0520-390:C2 & 05:22:48.3& -39:03:38 & S = ~68 & . . .~~\\
           &AM 0520-390:C3 & 05:22:44.3& -39:03:41 & S = ~23 & . . .~~\\
& & & & & \\
AM 0521-434&AM 0521-434:RN & 05:22:30.5& -43:46:17 & D = ~30 & 24,291\\
           &AM 0521-434:C1 & 05:22:30.4& -43:45:51 & S = ~26 & . . .~~\\
& & & & & \\
AM 0545-355&AM 0545-355:RN & 05:47:22.0& -35:49:29 & D = ~42 & 13,885\\
           &AM 0545-355:C1 & 05:47:19.4& -35:49:45 & S = ~36 & . . .~~\\
& & & & & \\
AM 0545-434&AM 0545-434:RN & 05:47:00.0& -43:44:56 & D = ~38 & . . .~~\\
           &AM 0545-434:C1 & 05:47:01.2& -43:44:33 & S = ~26 & . . .~~\\
& & & & & \\
AM 0642-645&AM 0642-645:RN & 06:42:34.2& -64:58:28 & D = ~20 & . . .~~\\
           &AM 0642-645:C1 & 06:42:32.7& -64:59:24 & S = ~05 & . . .~~\\
           &AM 0642-645:C2 & 06:42:33.4& -64:59:28 & S = ~05 & . . .~~\\
& & & & & \\
& & & & & \\
& & & & & \\
AM 0642-801&AM 0642-801:RN & 06:38:38.4& -80:14:50 & D = ~67 & 4,777\\
           &AM 0642-801:C1 & 06:38:36.6& -80:14:49 & S = ~08 & . . .~~\\
           &AM 0642-801:C2 & 06:38:27.9& -80:14:59 & S = ~29 & . . .~~\\
& & & & & \\
AM 0643-462&AM 0643-462:RN & 06:45:03.3& -46:26:56 & D = ~60 & 12,056\\
           &AM 0643-462:C1 & 06:45:00.2& -46:26:43 & S = ~36 & . . .~~\\
& & & & & \\
AM 0644-741&AM 0644-741:RN & 06:43:05.6& -74:14:12 & D = ~97 & 6,505\\
           &AM 0644-741:C1 & 06:43:06.1& -74:12:55 & S = ~78 & 7,000\\
           &AM 0644-741:C2 & 06:43:25.5& -74:15:27 & S = 110 & 6,750\\
           &AM 0644-741:C3 & 06:44:17.5& -74:16:37 & S = 326 & 6,430\\
& & & & & \\
AM 0755-785&AM 0755-785:RN & 07:53:31.5& -79:06:28 & D = ~27 & . . .~~\\
           &AM 0755-785:C1 & 07:53:35.6& -79:06:28 & S = ~13 & . . .~~\\
& & & & & \\
AM 0814-760&AM 0814-760:RN & 08:13:31.6& -76:17:29 & D = ~28 & . . .~~\\
           &AM 0814-760:C1 & 08:13:30.9& -76:17:41 & S = ~13 & . . .~~\\
& & & & & \\
& & & & & \\
& & & & & \\
AM 1003-215&AM 1003-215:RN & 10:05:43.2& -22:05:43 & D = ~28 & 26,329\\
           &AM 1003-215:C1 & 10:05:43.3& -22:05:53 & S = ~12 & . . .~~\\
           &AM 1003-215:C2 & 10:05:43.4& -22:05:03 & S = ~41 & . . .~~\\
           &AM 1003-215:C3 & 10:05:44.1& -22:06:19 & S = ~38 & . . .~~\\
           &AM 1003-215:C4 & 10:05:46.2& -22:05:52 & S = ~43 & . . .~~\\
& & & & & \\
AM 1006-380&AM 1006-380:RN & 10:09:05.2& -38:24:35 & D = ~80 & 4,943\\
           &AM 1006-380:C1 & 10:09:08.0& -38:23:48 & S = ~58 & 4,481\\
& & & & & \\
AM 1025-370&AM 1025-370:RN & 10:27:13.8& -37:25:17 & D = ~27 & . . .~~\\
           &AM 1025-370:C1 & 10:27:13.8& -37:25:33 & S = ~16 & . . .~~\\
& & & & & \\
AM 1133-245&AM 1133-245:RN & 11:35:30.8& -25:08:44 & D = ~67 & 11,704\\
           &AM 1133-245:C1 & 11:35:17.6& -25:08:45 & S = 180 & 11,682\\
           &AM 1133-245:C2 & 11:35:34.5& -25:05:41 & S = 189 & 11,601\\
& & & & & \\
AM 1135-284&AM 1135-284:RN & 11:37:42.2& -29:05:01 & D = ~24 & 5,558\\
           &AM 1135-284:C1 & 11:37:36.2& -29:04:11 & S = ~97 & . . .~~\\
& & & & & \\
& & & & & \\
AM 1152-421&AM 1152-241:RN & 11:55:21.7& -42:29:19 & D = ~46 & 4,617\\
           &AM 1152-241:C1 & 11:55:21.3& -42:29:17 & S = ~07 & . . .~~\\
& & & & \\
AM 1159-530&AM 1159-530:RN & 12:01:35.9& -53:21:28 & D = ~55 & . . .~~\\
           &AM 1159-530:C1 & 12:01:32.3& -53:20:47 & S = ~52 & . . .~~\\
& & & & \\
AM 1249-462&AM 1249-462:RN & 12:52:38.7& -46:38:59 & D = ~50 & . . .~~\\
           &AM 1249-462:C1 & 12:52:34.9& -46:40:04 & S = ~78 & . . .~~\\
& & & & & \\
AM 1251-283&AM 1251-283:RN & 12:54:42.4& -28:51:57 & D = ~29 & 17,003\\
           &AM 1251-283:C1 & 12:54:43.1& -28:52:17 & S = ~24 & 16,242\\
           &AM 1251-283:C2 & 12:54:43.1& -28:50:17 & S = 100 & 16,359\\
           &AM 1251-283:C3 & 12:54:36.5& -28:51:28 & S = ~82 & bckgrnd\\
           &AM 1251-283:C4 & 12:54:40.9& -28:53:03 & S = ~69 & 16,576\\
           &AM 1251-283:C5 & 12:54:49.7& -28:52:37 & S = 105 & foregrnd\\
& & & & & \\
AM 1300-412&AM 1300-412:RN & 13:03:00.3& -41:42:16 & D = ~18 & 3,422\\
           &AM 1300-412:C1 & 13:03:00.4& -41:42:20 & S = ~05 & . . .~~\\
& & & & & \\
& & & & & \\
AM 1308-253&AM 1308-253:RN & 13:11:08.7& -25:54:00 & D = ~44 & 13,556\\
           &AM 1308-253:C1 & 13:11:08.5& -25:54:00 & S = ~05 & . . .~~\\
           &AM 1308-253:C2 & 13:11:07.4& -25:54:28 & S = ~85 & . . .~~\\
& & & & & \\
AM 1323-222&AM 1323-222:RN & 13:26:20.3& -22:37:54 & D = ~88 & 4,627\\
           &AM 1323-222:C1 & 13:26:29.4& -22:33:49 & S = 272 & bckgrnd\\
& & & & & \\
AM 1325-251&AM 1325-251:RN & 13:27:58.1& -25:31:49 & D = ~18 & 10,936\\
           &AM 1325-251:C1 & 13:27:57.7& -25:31:44 & S = ~07 & . . .~~\\
           &AM 1325-251:C2 & 13:27:57.1& -25:30:50 & S = ~52 & . . .~~\\
& & & & & \\
AM 1354-250&AM 1354-250:RN & 13:57:13.7& -25:14:44 & D = ~53 & 6,150\\
           &AM 1354-250:C1 & 13:57:17.8& -25:13:30 & S = ~97 & 6,175\\
           &AM 1354-250:C2 & 13:57:24.6& -25:13:59 & S = 157 & . . .~~\\
& & & & & \\
AM 1358-221&AM 1358-221:RN & 14:01:08.0& -22:33:35 & D = ~53 & 10,940\\
           &AM 1358-221:C1 & 14:01:13.2& -22:32:45 & S = ~87 & . . .~~\\
           &AM 1358-221:C2 & 14:01:12.8& -22:31:52 & S = 121 & . . .~~\\
           &AM 1358-221:C3 & 14:01:25.6& -22:34:20 & S = 247 & 10,786\\
& & & & & \\
AM 1413-243&AM 1413-243:RN & 14:16:09.3& -24:50:30 & D = ~35 & 13,740\\
           &AM 1413-243:C1 & 14:16:04.9& -24:49:21 & S = ~94 & . . .~~\\
           &AM 1413-243:C2 & 14:16:09.3& -24:51:09 & S = ~37 & . . .~~\\
           &AM 1413-243:C3 & 14:16:13.3& -24:50:47 & S = ~56 & . . .~~\\
& & & & \\
AM 1425-234&AM 1425-234:RN & 14:55:09.8& -23:58:09 & D = ~61 & . . .~~\\
           &AM 1425-234:C1 & 14:55:09.4& -23:58:16 & S = ~09 & . . .~~\\
           &AM 1425-234:C2 & 14:55:08.0& -23:56:24 & S = 112 & . . .~~\\
           &AM 1425-234:C3 & 14:55:00.6& -23:55:44 & S = 195 & . . .~~\\
& & & & & \\
AM 1434-783&AM 1434-783:RN & 14:40:27.5& -78:48:35 & D = ~40 & 4,625\\
           &AM 1434-783:C1 & 14:40:32.2& -78:48:28 & S = ~40 & . . .~~\\
& & & & & \\
AM 1452-234&AM 1452-234:RN & 14:55:09.8& -23:58:09 & D = ~61 & 12,304\\
           &AM 1452-234:C1 & 14:55:09.4& -23:58:16 & S = ~09 & . . .~~\\
           &AM 1452-234:C2 & 14:55:08.0& -23:56:24 & S = 112 & . . .~~\\
           &AM 1452-234:C3 & 14:55:00.6& -23:55:44 & S = 195 & . . .~~\\
& & & & & \\
AM 1514-362&AM 1514-362:RN & 15:17:48.2& -36:34:56 & D = ~24 & . . .~~\\
           &AM 1514-362:C1 & 15:17:47.8& -36:35:00 & S = ~06 & 7,108\\

AM 1627-824&AM 1627-824:RN & 16:35:56.7& -82:50:17 & D = ~32 & . . .~~\\
           &AM 1627-824:C1 & 16:36:01.0& -82:50:12 & S = ~11 & . . .~~\\
           &AM 1627-824:C2 & 16:35:17.2& -82:51:17 & S = ~94 & . . .~~\\
& & & & & \\
AM 1724-622&AM 1724-622:RN & 17:29:09.6& -62:26:45 & D = 125 & 4,641\\
           &AM 1724-622:C1 & 17:29:25.3& -62:28:50 & S = 167 & 4,800\\
& & & & & \\
AM 1827-625&AM 1827-625:RN & 18:32:02.7& -62:55:44 & D = ~33 & 4,086\\
           &AM 1827-625:C1 & 18:32:02.6& -62:55:47 & S = ~04 & . . .~~\\
& & & & & \\
AM 1854-490&AM 1854-490:RN & 18:58:06.8& -49:00:41 & D = ~14 & 4,250\\
           &AM 1854-490:C1 & 18:58:05.5& -49:00:52 & S = ~18 & . . .~~\\
& & & & & \\
AM 1947-445&AM 1947-445:RN & 19:51:19.8& -44:52:40 & D = ~62 & 5,805\\
           &AM 1947-445:C1 & 19:51:26.3& -44:52:36 & S = 141 & 5,672\\
& & & & & \\
AM 1953-260&AM 1953-260:RN & 19:56:28.4& -25:54:56 & D = ~40 & . . .~~\\
           &AM 1953-260:C1 & 19:56:29.0& -25:55:16 & S = ~24 & 14,802 \\
& & & & & \\
& & & & & \\
AM 1957-394&AM 1957-394:RN & 20:00:36.0& -39:39:02 & D = ~51 & . . .~~\\
           &AM 1957-394:C1 & 20:00:35.4& -39:39:02 & S = ~05 & . . .~~\\
& & & & & \\
AM 2012-282&AM 2012-282:RN & 20:15:13.9& -28:18:25 & D = ~25 & 7,051\\
           &AM 2012-282:C1 & 20:15:13.6& -28:18:24 & S = ~06 & . . .~~\\
           &AM 2012-282:C2 & 20:15:14.2& -28:18:19 & S = ~08 & . . .~~\\
& & & & & \\
AM 2021-724&AM 2021-724:RN & 20:26:29.2& -72:36:46 & D = ~28 & . . .~~\\
           &AM 2021-724:C1 & 20:26:33.7& -72:36:08 & S = ~42 & 22,706\\
& & & & & \\
AM 2024-544&AM 2024-544:RN & 20:27:53.9& -54:38:00 & D = ~61 & 8,100\\
           &AM 2024-544:C1 & 20:27:57.0& -54:37:51 & S = ~28 & . . .~~\\
           &AM 2024-544:C2 & 20:27:48.5& -54:38:28 & S = ~55 & . . .~~\\
& & & & & \\
AM 2026-424&AM 2026-424:RN & 20:29:32.2& -42:30:24 & D = ~61 & . . .~~\\
           &AM 2026-424:C1 & 20:29:33.5& -42:30:24 & S = ~17 & 15,486\\
& & & & & \\
AM 2033-260&AM 2033-260:RN & 20:36:20.5& -25:56:52 & D = ~23 & 12,500\\
           &AM 2033-260:C1 & 20:36:23.9& -25:57:29 & S = ~60 & 12,257\\
           &AM 2033-260:C2 & 20:36:20.7& -25:56:41 & S = ~10 & . . .~~\\
AM 2034-483&AM 2034-483:RN & 20:38:08.7& -48:23:16 & D = ~22 & . . .~~\\
           &AM 2034-483:C1 & 20:38:06.2& -48:23:16 & S = ~26 & . . .~~\\
           &AM 2034-483:C2 & 20:38:02.8& -48:22:55 & S = ~63 & . . .~~\\
& & & & & \\
AM 2044-691&AM 2044-691:RN & 20:48:56.8& -69:05:32 & D = ~42 & 11,413\\
           &AM 2044-691:C1 & 20:48:56.6& -69:05:29 & S = ~03 & . . .~~\\
& & & & & \\
AM 2056-392&AM 2056-392:RN & 21:00:07.3& -39:17:57 & D = ~48 & . . .~~\\
           &AM 2056-392:C1 & 21:00:15.8& -39:20:05 & S = 150 & . . .~~\\
& & & & & \\
AM 2100-725&AM 2056-392:RN & 21:05:38.2& -72:47:09 & D = ~27 & . . .~~\\
           &AM 2056-392:C1 & 21:05:54.5& -72:47:20 & S = ~72 & 20,984\\
           &AM 2056-392:C2 & 21:05:52.9& -72:47:22 & S = ~65 & . . .~~\\
           &AM 2056-392:C3 & 21:05:49.5& -72:46:21 & S = ~69 & . . .~~\\
           &AM 2056-392:C4 & 21:05:58.8& -72:47:59 & S = 103 & . . .~~\\
           &AM 2056-392:C5 & 21:05:16.2& -72:47:58 & S = 109 & . . .~~\\
           &AM 2056-392:C6 & 21:05:11.5& -72:47:04 & S = 119 & . . .~~\\
           &AM 2056-392:C7 & 21:05:48.2& -72:47:29 & S = ~48 & . . .~~\\
& & & & & \\
& & & & & \\
AM 2107-474&AM 2107-472:RN & 21:10:31.0& -47:30:32 & D = ~31 & 5,060\\
           &AM 2107-472:C1 & 21:10:31.4& -47:30:36 & S = ~05 & . . .~~\\
& & & & & \\
AM 2128-302&AM 2128-302:RN & 21:31:08.2& -30:16:27 & D = ~84 & 10,037\\
           &AM 2128-302:C1 & 21:31:01.5& -30:19:40 & S = 208 & 9,199\\
& & & & & \\
AM 2131-254&AM 2131-254:RN & 21:34:29.2& -25:28:38 & D = ~33 & 16,218\\
           &AM 2131-254:C1 & 21:34:31.1& -25:28:46 & S = ~29 & . . .~~\\
& & & & & \\
AM 2131-495&AM 2131-495:RN & 21:34:37.6& -49:42:46 & D = ~27 & 21,645\\
           &AM 2131-495:C1 & 21:34:37.2& -49:42:32 & S = ~15 & . . .~~\\
& & & & & \\
AM 2132-535&AM 2132-535:RN & 21:36:11.1& -53:41:26 & D = ~32 & . . .~~\\
           &AM 2132-535:C1 & 21:36:13.0& -53:42:54 & S = ~91 & . . .~~\\
           &AM 2132-535:C2 & 21:36:27.1& -53:43:09 & S = 176 & . . .~~\\
           &AM 2132-535:C3 & 21:36:26.7& -53:43:31 & S = 188 & . . .~~\\
& & & & & \\
AM 2134-471&AM 2134-471:RN & 21:37:28.0& -47:02:09 & D = ~52 & 9,254\\
           &AM 2134-471:C1 & 21:37:31.1& -47:00:38 & S = ~98 & . . .~~\\
           &AM 2134-471:C2 & 21:37:35.9& -47:02:40 & S = ~90 & . . .~~\\

AM 2136-492&AM 2136-492:RN & 21:39:25.6& -49:09:36 & D = 40/72 & 15,990\\
           &AM 2136-492:C1 & 21:39:31.6& -49:08:11 & S = 104 & . . .~~\\
& & & & & \\
AM 2141-515&AM 2141-515:RN & 21:45:15.1& -51:41:54 & D = ~33 & . . .~~\\
           &AM 2141-515:C1 & 21:45:14.9& -51:41:49 & S = ~05 & . . .~~\\
& & & & & \\
AM 2145-543&AM 2145-543:RN & 21:49:08.5& -54:18:14 & D = ~22 & 19,544\\
           &AM 2145-543:C1 & 21:49:08.2& -54:17:53 & S = ~22 & . . .~~\\
& & & & & \\
AM 2152-592&AM 2152-592:RN & 21:56:19.4& -59:07:15 & D = ~15 & 22,351\\
           &AM 2152-592:C1 & 21:56:17.5& -59:07:17 & S = ~15 & . . .~~\\
           &AM 2152-592:C2 & 21:56:20.8& -59:06:51 & S = ~27 & . . .~~\\
& & & & & \\
AM 2155-263&AM 2155-263:RN & 21:58:38.4& -26:22:41 & D = ~27 &  21,255\\
           &AM 2155-263:C1 & 21:58:37.2& -26:22:25 & S = ~23 & . . .~~\\
& & & & & \\
& & & & & \\
& & & & & \\
& & & & & \\
& & & & & \\
AM 2200-715&AM 2200-715:RN & 22:04:57.5& -71:42:14 & D = ~32 & . . .~~\\
           &AM 2200-715:C1 & 22:04:22.4& -71:43:20 & S = 177 & 9,350\\
           &AM 2200-715:C2 & 22:04:12.7& -71:35:08 & S = 478 & . . .~~\\
           &AM 2200-715:C3 & 22:04:47.8& -71:35:59 & S = 379 & . . .~~\\
           &AM 2200-715:C4 & 22:05:51.5& -71:39:51 & S = 288 & . . .~~\\
           &AM 2200-715:C5 & 22:06:23.9& -71:43:53 & S = 418 & . . .~~\\
           &AM 2200-715:C6 & 22:05:39.0& -71:46:45 & S = 330 & . . .~~\\
& & & & & \\
AM 2201-230&AM 2201-230:RN & 22:04:19.3& -22:47:33 & D = ~24 & 21,732\\
           &AM 2201-230:C1 & 22:04:19.5& -22:47:48 & S = ~17 & 21,675\\
& & & & & \\
AM 2220-493&AM 2220-493:RN & 22:23:13.4& -49:17:46 & D = ~32 & . . .~~\\
           &AM 2220-493:C1 & 22:23:12.7& -49:17:16 & S = ~31 & 17,947\\
& & & & & \\
AM 2230-481&AM 2230-481:RN & 22:33:47.0& -48:01:30 & D = ~52 & 10,476\\
           &AM 2230-481:C1 & 22:33:43.3& -48:01:31 & S = ~37 & 10,361\\
& & & & & \\
AM 2238-541&AM 2238-541:RN & 22:41:55.1& -53:58:43 & D = ~32 & . . .~~\\
           &AM 2238-541:C1 & 22:41:57.5& -53:59:45 & S = ~67 & . . .~~\\
& & & & & \\
AM 2240-304&AM 2240-304:RN & 22:43:36.9& -30:28:50 & D = ~22 & 17,449\\
           &AM 2240-304:C1 & 22:43:36.3& -30:28:55 & S = ~08 & . . .~~\\
& & & & & \\
AM 2302-322&AM 2302-322:RN & 23:05:39.9& -32:09:07 & D = ~25 & 18,137\\
           &AM 2302-322:C1 & 23:05:41.9& -32:08:11 & S = ~66 & 18,077\\
           &AM 2302-322:C2 & 23:05:43.9& -32:09:46 & S = ~59 & bckgrnd\\
& & & & & \\
AM 2308-324&AM 2308-324:RN & 23:11:17.3& -32:27:07 & D = ~33 & 11,327\\
           &AM 2308-324:C1 & 23:11:16.9& -32:27:21 & S = ~15 & . . .~~\\
& & & & & \\
AM 2313-261&AM 2313-261:RN & 23:15:46.5& -25:54:20 & D = ~20 & 6,472\\
           &AM 2313-261:C1 & 23:15:46.5& -25:54:28 & S = ~09 & . . .~~\\
& & & & & \\
AM 2317-672&AM 2317-672:RN & 23:20:09.7& -67:08:58 & D = ~32 & . . .~~\\
           &AM 2317-672:C1 & 23:20:11.2& -67:09:00 & S = ~10 & . . .~~\\
           &AM 2317-672:C2 & 23:20:11.6& -67:08:28 & S = ~33 & . . .~~\\
& & & & & \\
AM 2322-671&AM 2322-671:RN & 23:25:42.8& -67:03:29 & D = ~19 & . . .~\\
           &AM 2322-671:C1 & 23:25:45.3& -67:03:33 & S = ~15 & . . .~\\
& & & & & \\
AM 2323-512&AM 2323-512:RN & 23:26:31.8& -51:08:07 & D = ~21 & . . .~\\
           &AM 2323-512:C1 & 23:26:31.2& -51:07:45 & S = ~23 & . . .~\\
& & & & & \\
AM 2338-312&AM 2338-312:RN & 23:41:26.0& -31:03:22 & D = ~54 & 17,538\\
           &AM 2338-312:C1 & 23:41:19.2& -31:03:01 & S = ~91 & 17,268\\
& & & & & \\
AM 2343-703&AM 2343-703:RN & 23:46:39.3& -70:22:51 & D = ~44 & . . .~\\
           &AM 2343-703:C1 & 23:46:38.5& -70:22:25 & S = ~27 & . . .~\\
& & & & & \\
AM 2353-291&AM 2353-291:RN & 23:56:23.8& -29:01:25 & D = ~36 & 8,931\\
           &AM 2353-291:C1 & 23:56:25.1& -29:01:24 & S = ~17 & 8,953\\

\enddata
\end{deluxetable}

\begin{deluxetable}{llcclr}
\tablecolumns{5}
\tablewidth{6.6truein}
\tablehead{
\colhead{Host Ring System}  & \colhead{Component}  & \colhead{RA (2000)}    & \colhead{Dec (2000)}  & \colhead{~~D/S} &\colhead{V$_{\odot}$}\\
\colhead{}    & \colhead{Name}    & \colhead{(h:m:s)} & \colhead{(d:m:s)}   & \colhead{~~(arcsec)} &\colhead{(km/s)} 
}

\startdata
ARP 146&ARP 146:RN & 00:06:44.3& -06:38:08 & D = ~19 & 22,616\\
       &ARP 146:C1 & 00:06:44.7& -06:38:13 & D = ~09 & . . .~~\\
& & & & & \\
ARP 318&ARP 318:RN & 02:09:24.4& -10:08:10 & D = ~40 & 4,073\\
       &ARP 318:C1 & 02:09:20.7& -10:08:00 & S = ~55 & 3,864\\
       &ARP 318:C2 & 02:09:38.3& -10:08:48 & S = 208 & 3,851\\
       &ARP 318:C3 & 02:09:42.7& -10:11:02 & S = 319 & 3,934\\
& & & & & \\
ARP 010&ARP 010:RN & 02:18:26.2& +05:39:14 & D = ~44 & 9,108\\
       &ARP 010:C1 & 02:18:28.6& +05:40:08 & S = ~65 & . . .~~\\
& & & & & \\
ARP 273&ARP 273:RN & 02:21:28.6& +39:22:32 & D = 100 & 7,563\\
       &ARP 273:C1 & 02:21:32.7& +39:21:24 & S = ~82 & 7,335\\
& & & & & \\
ARP 145&ARP 145:RN & 02:23:07.7& +41:21:11 & D = ~57 & 5,425\\
       &ARP 145:C1 & 02:23:11.1& +41:22:04 & S = ~36 & 5,852\\
& & & & & \\
NGC 0985&NGC 0985:RN & 02:34:37.6& -08:47:15 & D = ~36 & 13,168\\
        &NGC 0985:C1 & 02:34:38.6& -08:48:03 & S = ~47 & . . .~~\\
& & & & & \\
ARP 118&ARP 118:RN & 02:55:12.0& -00:11:03 & D = ~30 & 8,648\\
       &ARP 118:C1 & 02:55:09.5& -00:10:40 & S = ~45 & 8,459\\
       &ARP 118:C2 & 02:55:06.5& -00:09:46 & S = 114 & 8,417\\
& & & & & \\
ARP 147&ARP 147:RN & 03:11:18.6& +01:18:51 & D = ~17 & 9,656\\
       &ARP 147:C1 & 03:11:19.4& +01:18:48 & S = ~14 & 9,415\\
& & & & & \\
IC 1908&IC 1908:RN & 03:15:05.4& -54:49:09 & D = ~30 & 8,234\\
       &IC 1908:C1 & 03:15:05.7& -54:49:20 & S = ~13 & . . .~~\\
& & & & & \\
ESO 200-IG 009&ESO 200-IG 009:RN & 03:21:40.5& -51:39:31 & D = ~29 & 17,328\\
              &ESO 200-IG 009:C1 & 03:21:41.0& -51:39:49 & S = ~21 & 20,980\\
              &ESO 200-IG 009:C2 & 03:21:43.6& -51:39:42 & S = ~32 & . . .~~\\
& & & & & \\
ARP 219&ARP 219:RN & 03:39:53.2& -02:06:47 & D = ~46 & 10,488\\
       &ARP 219:C1 & 03:39:54.8& -02:07:25 & S = ~44 & . . .~~\\
& & & & & \\
ARP 141&ARP 141:RN & 07:14:20.0& +73:28:25 & D = ~75 & 2,728\\
       &ARP 141:C1 & 07:14:20.2& +73:28:51 & S = ~26 & 2,735\\
& & & & & \\
ARP 143&ARP 143:RN & 07:46:55.0& +39:00:56 & D = ~87 & 4,257\\
       &ARP 143:C1 & 07:46:52.9& +39:01:55 & D = ~65 & 4,048\\
       &ARP 143:C2 & 07:46:55.5& +39:00:27 & D = ~30 & 4,002\\
& & & & & \\
NGC 2793&NGC 2793:RN & 09:16:47.1& +34:25:56 & D = ~46 & . . .~~\\
        &NGC 2793:C1 & 09:16:47.2& +34:25:48 & S = ~11 & 1,687\\
        &NGC 2793:C2 & 09:16:40.8& +34:26:52 & S = ~97 & bckgrnd\\
        &NGC 2793:C3 & 09:16:46.6& +34:26:14 & S = ~18 & 1,667\\
& & & & & \\
ARP 142&ARP 142:RN & 09:37:44.1& +02:45:39 & D = ~53 & 7,225\\
       &ARP 142:C1 & 09:37:45.0& +02:44:51 & S = ~52 & 6,806\\
       &ARP 142:C2 & 09:37:41.2& +02:46:45 & S = ~81 & . . .~~  \\
& & & & & \\
IC 0614&IC 0614:RN & 10:26:51.8& -03:27:53 & D = ~34 & 10,258\\
       &IC 0614:C1 & 10:26:51.9& -03:27:39 & S = ~16 & . . .~~\\
       &IC 0614:C2 & 10:26:49.0& -03:29:19 & S = ~97 & 18,097\\
& & & & & \\
ARP 107&ARP 107:RN & 10:52:14.8& +30:03:28 & D = ~55 & 10,372\\
       &ARP 107:C1 & 10:52:18.4& +30:04:20 & D = ~70 & 10,665\\
& & & & & \\
ARP 148&ARP 148:RN & 11:03:52.4& +40:50:54 & D = ~18 & 10,551\\
       &ARP 148:C1 & 11:03:53.8& +40:51:00 & S = ~17 & . . .~~\\
& & & & & \\
ARP 335&ARP 335:RN & 11:04:23.5& +04:49:44 & D = 130 & 7,738\\
       &ARP 335:C1 & 11:04:23.5& +04:49:29 & S = ~15 & 7,763\\
       &ARP 335:C2 & 11:04:16.5& +04:52:10 & S = 180 & 7,647\\
& & & & & \\
VII Zw 466&VII Zw 466:RN & 12:32:04.9& +66:24:13 & D = ~21 & 14,490\\
          &VII Zw 466:C1 & 12:32:10.6& +66:24:19 & S = ~34 & . . .~~\\
          &VII Zw 466:C2 & 12:32:13.1& +66:23:59 & S = ~50 & 14,100\\
          &VII Zw 466:C3 & 12:32:11.6& +66:23:21 & S = ~65 & 14,360\\
& & & & & \\
NGC 4774&NGC 4774:RN & 12:53:06.4& +36:49:10 & D = ~23 & 8,373\\
        &NGC 4774:C1 & 12:53:05.8& +36:49:34 & S = ~25 & 8,373\\
& & & & & \\
VV 256&VV 256:RN & 14:00:54.4& +40:59:19 & D = ~64 & 3,738\\
      &VV 256:C1 & 14:00:56.2& +41:00:22 & S = ~68 & 3,834\\
& & & & & \\
KIG~0686&KIG~0686:RN & 15:32:56.9& +46:27:07 & D = ~11 & . . .~~\\
      &KIG~0686:C1 & 15:32:57.5& +46:27:14 & S = ~09 &~656\\
& & & & & \\
KIG~0700&KIG~0700:RN & 15:45:14.8& +00:46:24 & D = ~19 & . . .~~\\
      &KIG~0700:C1 & 15:45:14.3& +00:46:21 & S = ~08 & 3,839\\
& & & & & \\
ARP 125&ARP 125:RN & 16:38:13.9& +41:56:19 & D = ~45 & 8,572\\
       &ARP 125:C1 & 16:38:13.6& +41:55:51 & S = ~31 & 8,404\\
& & & & & \\
ARP 150&ARP 150:RN & 23:19:30.0& +09:30:30 & D = ~34 & 11,879\\
       &ARP 150:C1 & 23:19:31.0& +09:30:11 & S = ~26 & 11,562\\
       &ARP 150:C2 & 23:19:33.7& +09:29:42 & S = ~75 & 11,637\\
       &ARP 150:C3 & 23:19:27.7& +09:29:40 & S = ~59 & 12,350\\
& & & & & \\
ARP 284&ARP 284:RN & 23:36:14.2& +02:09:17 & D = ~44 & 2,798\\
       &ARP 284:C1 & 23:36:21.8& +02:09:25 & S = 111 & 2,771\\
\enddata
\end{deluxetable}

 \end{document}